\documentclass{article}

\usepackage{paper_style}

\usepackage{graphicx}
\usepackage{amsmath}
\usepackage{amssymb}
\usepackage{booktabs}
\usepackage{multirow}
\usepackage{multicol}
\usepackage{xcolor}
\usepackage{colortbl}
\usepackage{appendix}
\usepackage{tcolorbox}
\usepackage{subcaption}
\usepackage{wrapfig}
\usepackage[pagebackref,breaklinks,colorlinks]{hyperref}

\usepackage[capitalize]{cleveref}
\crefname{section}{Sec.}{Secs.}
\Crefname{section}{Section}{Sections}
\Crefname{table}{Table}{Tables}
\crefname{table}{Tab.}{Tabs.}

\newcommand{\subfour}[1]{\vspace*{3mm}{\noindent\bf #1}}

\title{SVLA: A Unified Speech-Vision-Language Assistant with Multimodal Reasoning and Speech Generation}

\author{
  Ngoc Dung Huynh \\
  Deakin University \\
  Geelong, Australia\\
  \texttt{ndhuynh@deakin.edu} \\
   \And
  Mohamed Reda Bouadjenek \\
  Deakin University \\
  Geelong, Australia\\
  \texttt{reda.bouadjenek@deakin.edu.au} \\
  \And
  Imran Razzak \\
  Mohamed bin Zayed University of Artificial Intelligence \\
  Abu Dhabi, UAE\\
  \texttt{imran.razzak@mbzuai.ac.ae} \\
  \And
  Hakim Hacid \\
  Technology Innovation Institute (TII) \\
  Abu Dhabi, UAE\\
  \texttt{hakim.hacid@tii.ae} \\
  \And
  Sunil Aryal \\
  Deakin University \\
  Geelong, Australia\\
  \texttt{sunil.aryal@deakin.edu.au} \\
}

\begin{document}
\maketitle

\begin{abstract}
Large Vision-Language Models have shown impressive capabilities in tasks such as image captioning, visual question answering, and cross-modal retrieval.  
However, there are still  significant challenges that need to be addressed in order to fully unlock the potential of these models.
First, integrating speech, text, and vision into a unified model is particularly difficult for tasks like Spoken Image Captioning and Spoken Visual Question Answering, where the interaction between these modalities introduces additional complexity. 
Second, existing speech generation approaches differ—some generate speech directly, while others use an intermediate text step—but their impact on fluency, coherence, and accuracy remains unexplored. 
To address these challenges, we propose \textbf{SVLA}, a unified \textbf{S}peech-\textbf{V}ision-\textbf{L}anguage \textbf{A}ssistant based on a decoder-only transformer architecture that seamlessly integrates multimodal inputs and outputs. 
We enhance model performance with a large-scale speech-text-image dataset containing 38.2 million examples and 64.1 hours of TTS-generated speech. 
Our approach advances multimodal understanding and generation, facilitating more effective integration of speech, text, and vision (\href{http://github.com/vlm-svla/svla}{http://github.com/vlm-svla/svla}). 
\end{abstract}

\section{Introduction}

Recent advances in Large Vision-Language Models (LVLMs)~ \cite{openai2023,li2023blip,liu2024visual,alayrac2022flamingo} mark a significant step toward multimodal AI, enabling models to interpret and reason over visual and textual inputs. However, these models remain constrained by their reliance on text-based instructions and lack native support for speech.
Efforts such as LLaMA 3~ \cite{touvron2023llama} and Qwen-Audio~ \cite{chu2023qwen} have begun incorporating speech through modality-specific encoders, these models remain limited to speech perception and do not support speech generation.

Recent models~\cite{zhang2023speechgpt,zhan2024anygpt,wu2024next,xu2025qwen2} address this limitation by introducing discrete speech representations, allowing language models to process speech as semantic tokens within a unified speech-text token space. This enables bidirectional modeling and multimodal integration. However, most systems remain focused on shallow tasks—such as text-to-speech~\cite{Veaux2017CSTRVC}, image-to-music~\cite{chowdhury2024melfusion}, or basic audio captioning~\cite{kim2019audiocaps}—and struggle with more complex reasoning tasks like image captioning (IC) or visual question answering (VQA) involving spoken inputs or outputs.
A key bottleneck is the absence of large-scale, richly aligned datasets spanning speech, text, and vision. Most models are trained on unimodal or bimodal data, limiting their ability to generalize to cognitively demanding, speech-enabled multimodal reasoning.

Furthermore, two prominent paradigms have emerged for enabling speech capabilities within multimodal systems \cite{zhang2023speechgpt,nachmani2023spoken}: Cross-modal Instruction and Chain-of-Modality Instruction. Cross-modal Instruction directly maps inputs across modalities, such as converting an image to speech, offering efficiency but frequently compromising semantic coherence. In contrast, Chain-of-Modality Instruction involves first translating the input into textual form, performing reasoning over the generated text, and subsequently producing output in speech or text form. This method leverages the structured reasoning and planning capabilities inherent to large language models but may lead to verbosity. However, a comparative analysis of these paradigms under consistent tri-modal settings within LVLMs has not yet been conducted.

To address the above challenges, we introduce a large-scale synthetic trimodal dataset—combining text, image, and speech—to support instruction-following in both written and spoken formats. This dataset integrates widely-used resources, including VQAv2 \cite{goyal2017making}, LAION-600M \cite{laion}, Visual Genome \cite{gva}, LibriHeavy \cite{kang2024libriheavy}, LibriSpeech \cite{panayotov2015librispeech}, CommonVoice \cite{ardila2019common}, A-OKVQA \cite{schwenk2022okvqa}, VizWiz \cite{gurari2018vizwiz}, GQA \cite{hudson2019gqa}, and COCO-Caption \cite{lin2014microsoft}.
To ensure natural-sounding and diverse speech data, we use a controllable text-to-speech (TTS) system that varies accent, speaking speed, and prosody. We further include noise and stylistic augmentation from MUSAN \cite{snyder2015musan} to improve robustness. Additionally, we implement a modality-switching instruction tuning strategy, enabling the model to flexibly process and respond in either text or speech.

Building on this foundation, we propose SVLA (Speech-Vision-Language Assistant), a unified, self-supervised multimodal model capable of reasoning over image, text, and speech inputs and generating outputs in either text or speech. SVLA features a hybrid fusion architecture: it applies early fusion between speech and text using discrete semantic units, allowing both to be modeled jointly in a shared language space~ \cite{zhan2024anygpt, zhang2023speechtokenizer}, and late fusion for visual input by integrating image embeddings from a pretrained encoder~ \cite{liu2024visual}. This design supports complex tasks such as spoken VQA, speech-driven image captioning, and multimodal instruction following.

We also present the systematic evaluation comparing Cross-modal and Chain-of-Modality Instruction within a unified vision-language-speech framework. Our experiments span four controlled configurations—text-to-text, text-to-speech, speech-to-text, and speech-to-speech—using consistent instruction formats and shared example naming across tasks. Benchmarks such as VQAv2 and COCO-Captions are used to assess the trade-offs between fluency, coherence, and semantic quality under each setting.

Our contributions are as follows:
\begin{itemize}
    \item We construct a large-scale, aligned speech-text-vision dataset from diverse benchmarks, supporting instruction-following in both textual and spoken forms.
    \item We propose SVLA, a unified tri-modal model that integrates early fusion (text-speech) and late fusion (vision) to enable robust multimodal reasoning and generation. 
    \item We establish an evaluation framework that systematically compares Cross-modal and Chain-of-Modality paradigms across consistent input-output modality configurations, while also assessing robustness to variations in accent and speaking speed.
\end{itemize}

\section{Related Works}

\subfour{Speech-Enabled Vision-Language Models:}
LVLMs \cite{yuan2021florence,li2023blip,liu2024visual,chen2024internvl} were initially designed for image-text reasoning, lacking native speech support. Early extensions added speech input and TTS output but remained limited in conversational expressiveness \cite{openai2023gpt4,dubey2024llama}, with higher latency and limited contextual adaptability due to multi-stage pipelines. GPT-4o \cite{chatgpt4o}addresses these issues with native speech generation for real-time, expressive interaction, though its closed-source nature has driven open efforts to replicate its capabilities. Models like NExT-GPT \cite{wu2024next}, CoDi-1/2 \cite{tang2023any,tang2024codi}, Unified-IO \cite{lu2024unified}, and AnyGPT \cite{zhan2024anygpt} extend multimodal support via modality-specific encoders or projections, targeting perceptual tasks (e.g., image/audio/video generation, ASR, TTS). However, cross-modal reasoning (e.g., VQA, image captioning across text and speech modalities) remains underexplored. TMT \cite{kim2024tmt} offers limited progress, enabling speech output from image captions but lacking joint multimodal reasoning.

 \subfour{Speech-Text-Vision Datasets:}
Most speech datasets used in multimodal learning—such as LibriSpeech \cite{panayotov2015librispeech}, CommonVoice \cite{ardila2019common}, and GigaSpeech \cite{chen2021gigaspeech}—are designed for ASR or TTS and do not include vision. While datasets like SpeechCOCO \cite{havard2017speech} and SpokenCOCO~ \cite{hsu2020text} add speech to image captioning, they remain limited in scale, task diversity, and interactivity. Others, like How2 \cite{sanabria2018how2}, offer aligned speech and video, but focus on narrow instructional domains and do not support flexible input-output modality switching. Some recent works synthesize multimodal dialogues by using large language models to generate text-based prompts, which are paired with audio and visual content using tools like TTS and image generation models. While useful for data augmentation, these datasets often focus on perception tasks and lack cognitively demanding reasoning challenges such as VQA or multimodal instruction following.

\subfour{Direct vs. Instruction-Based Speech Generation:} 
Speech generation in multi-modal systems generally follows one of two paradigms: direct generation, where speech is produced end-to-end without textual intermediates, and instruction-based generation, where models first generate text and then synthesize speech from it. Instruction-based approaches—used in models like GPT-4o~\cite{chatgpt4o}, AnyGPT~\cite{zhan2024anygpt}, NExT-GPT~\cite{wu2024next}, and SpeechGPT~\cite{zhang2023speechgpt}—offer modality consistency but often produce redundant or unnatural phrasing. In contrast, direct approaches aim to generate speech directly from inputs (e.g., Spectron~\cite{nachmani2023spoken}), allowing for more fluid prosody and conversational tone.

\section{Data Generation}
\begin{figure}[t]
    \centering
    \includegraphics[width=0.8\linewidth]{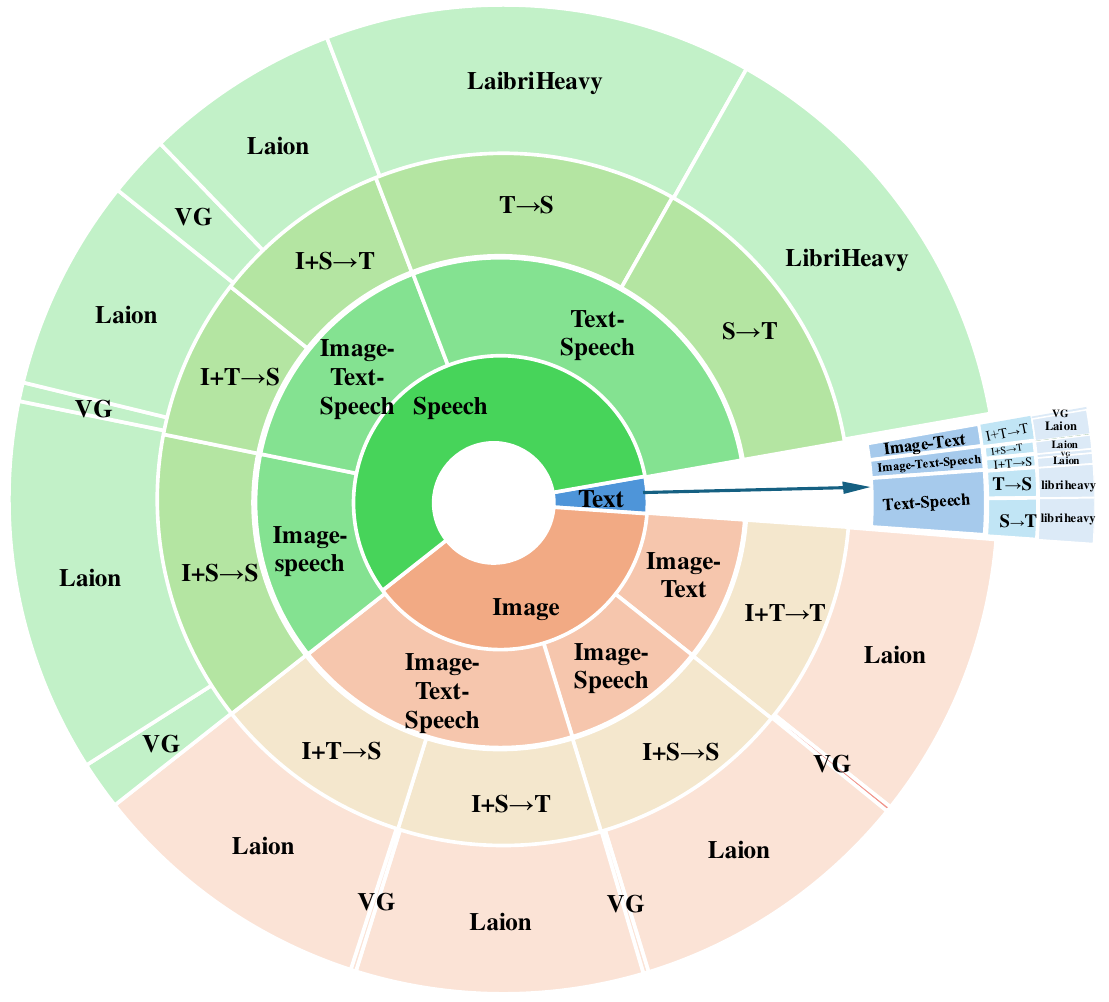}
    \caption{Pre-training data distribution across three modalities: Image, Text, and Speech.}
    \label{fig:pre-training-pie-chart}
\end{figure}
To construct a large-scale tri-modal dataset, we extend existing image-text corpora~\cite{lin2014microsoft, laion} with corresponding spoken utterances. These datasets are selected for their abundance, diversity, and broad coverage of real-world visual and linguistic contexts, providing a strong foundation for scalable multimodal learning. We synthesize speech from text using the Melon-TTS model~\cite{zhao2024melo}, chosen for its controllability over prosody and speaker characteristics. To enhance diversity, we vary speaking rates from 0.7× to 1.3× and include multiple English accents, such as American, British, Indian, and Australian. Additionally, we introduce environmental background noise—e.g., rain, footsteps, and ambient sounds—from the MUSAN corpus~\cite{snyder2015musan} to simulate realistic acoustic conditions. All audio is standardized to a 16 kHz sampling rate to ensure clarity and computational efficiency. This pipeline yields a rich and varied speech-text-image dataset suitable for both pre-training and fine-tuning multimodal models.

\subsection{Pre-train dataset}

Our pre-training strategy targets a unified speech-vision-language model capable of TTS, ASR, image captioning, and VQA. The training data includes: (1) 8M text-speech pairs from the publicly available LibriHeavy corpus~\cite{kang2024libriheavy} for speech generation and recognition, and (2) 6M image-text-speech triples, where speech is \textit{synthetically generated} from image-caption pairs in LAION-COCO~\cite{laion} and question-answer pairs from Visual Genome~\cite{gva} to support vision-language reasoning.
To enable modality switching, we use instruction-style prompts (e.g., “Answer this question in speech” or “Describe the image in text”), guiding the model to produce either spoken or written outputs. Speech outputs are capped at 5 seconds for training efficiency. Figure~\ref{fig:pre-training-pie-chart} shows the modality token distribution\footnote{Token counts: text via Qwen-2.5-1.5B~\cite{yang2024qwen2}, speech via SpeechTokenizer~\cite{zhang2023speechtokenizer}, and images estimated at 256 tokens each.}. Additional details are provided in Appendix~\ref{sec:appendix_data_generation}.

\subsection{Visual Instruction dataset}

For supervised fine-tuning (SFT), we adopt a similar structure to pre-training but introduce multi-turn conversations to improve coherence and long-range context retention. The text-speech subset includes LibriSpeech~\cite{panayotov2015librispeech} and CommonVoice~\cite{ardila2019common}, covering diverse linguistic and command-oriented expressions. The vision-text-speech subset incorporates VQAv2, A-OKVQA~\cite{schwenk2022okvqa}, GQA~\cite{hudson2019gqa}, VizWiz~\cite{gurari2018vizwiz}, and COCO-Captions-2014, ensuring broad coverage of open-ended and visually grounded tasks.
To enable consistent comparison across modalities, we design all tasks to support four input-output configurations: text-to-text, text-to-speech, speech-to-text, and speech-to-speech. Dialogue examples are constructed to fit within a 10-second speech limit and are paired with instruction prompts that guide the model to transition between text, speech, and vision as required. Additional details on the SFT dataset construction are provided in Appendix~\ref{sec:appendix_data_generation}.

\section{Model Architecture}

\begin{figure*}[h]
    \centering
    \includegraphics[width=0.9\linewidth]{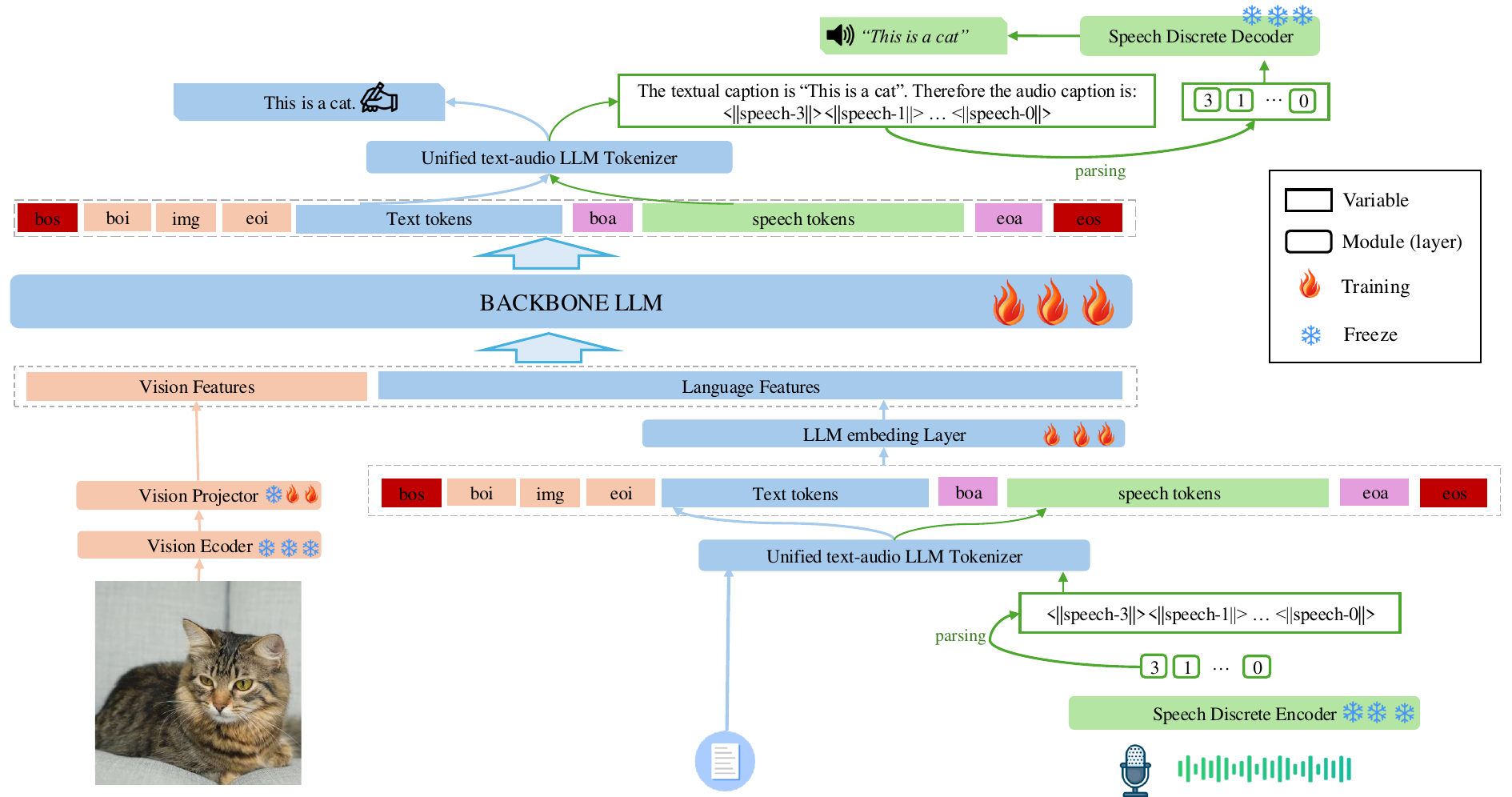}
    \caption{SVLA Architecture}
    \label{fig:arch}
\end{figure*}

Our architecture is a hybrid of LLaVA~\cite{liu2024visual} and AnyGPT~\cite{zhan2024anygpt}, combining visual grounding with speech processing via discrete tokens. It integrates speech and vision inputs into a unified language model, denoted as $f_{\theta}(x)$, where $\theta$ represents model parameters. The overall architecture is shown in Figure~\ref{fig:arch}.

\subsection{Vision Encoder}  

Given an input image $X_v$, we employ a pretrained VIT-based vision encoder, such as CLIP \cite{radford2021learning}, denoted as  $g(\cdot)$, to extract high-level visual representations:
\begin{equation}
Z_v = g(X_v), \quad Z_v \in \mathbb{R}^{n_v \times d_v}.
\end{equation}
Here, $Z_v$ is obtained from the final layer of the vision encoder, where $n_v$ is the number of image patches. Each visual token in $Z_v$ has a feature dimension of $d_v$.  

Next, to align visual embeddings with the LLM’s word embedding space, we apply a learnable linear projection layer $W_v$:
\begin{equation}
H_v = W_v Z_v, \quad H_v \in \mathbb{R}^{n_v \times d_h}.
\end{equation}
Here, $H_v$ represents the projected visual tokens, now residing in the same dimensional space $d_h$ as the LLM’s text embeddings, ensuring compatibility for multimodal fusion.

\subsection{Speech Encoder and Tokenization}

For speech input \( X_s \), a pretrained speech encoder \( s(\cdot) \) generates a sequence of discrete tokens:
\[
Z_s = \{ q_1, q_2, \dots, q_{T_s} \}, \quad q_i \in \{1, 2, \dots, d_s\},
\]
where \( T_s \) is the number of tokens and \( d_s \) is the speech vocabulary size. Each token \( q_i \) is mapped to a text-like representation (e.g., \(\langle\!\langle \text{speech\--}i\rangle\!\rangle\)) and added to the LLM’s vocabulary, enabling text and speech to be handled uniformly by the tokenizer.

\subsection{Multimodal Fusion}

Given text tokens $Z_t$, speech tokens $Z_s$, and visual embeddings $H_v$, we construct the multimodal input sequence:
\[
X = [\texttt{bos},\, \texttt{boi},\, \texttt{img},\, \texttt{eoi},\, Z_t,\, \texttt{boa},\, Z_s,\, \texttt{eoa},\, \texttt{eos}],
\]
where $n = |X|$ is the sequence length. Special tokens mark modality boundaries and control flow.

After embedding, we replace the placeholder token \texttt{img} with $H_v$, yielding:
\[
\begin{aligned}
E' = [&E_{\texttt{bos}},\, E_{\texttt{boi}},\, H_v,\, E_{\texttt{eoi}},\, E_{Z_t}, \\
      &E_{\texttt{boa}},\, E_{Z_s},\, E_{\texttt{eoa}},\, E_{\texttt{eos}}],
\end{aligned}
\]
with $E' \in \mathbb{R}^{n' \times d_h}$, where $n' = n_v + n - 1$. The LLM then processes $E'$ for multimodal understanding.

\subsection{Speech Decoding}

The multimodal LLM outputs a sequence of predicted tokens, from which the speech-related tokens \(Z_s'\) are extracted between the special tokens \(boa\) and \(eoa\). A pretrained speech decoder \(s^{-1}(\cdot)\) then transforms these discrete tokens back into speech waveforms:
$$
X_s' = s^{-1}(Z_s'),
$$
where \(Z_s'\) represents the predicted discrete speech token sequence, and \(X_s'\) is the resulting synthesized speech audio.

\subsection{Instruction-Based Speech Generation}
During pre-training (see Appendix~\ref{subsec:training_stages}), the model learns to generate both text and speech tokens directly. In SFT, text responses are always generated directly, while speech outputs follow two alternative strategies: direct generation or instruction-based generation, which we compare experimentally.
For simpler tasks like ASR and TTS, we apply lightweight prompts (e.g., \textit{“The transcript of the given speech is:”} for ASR, and \textit{“This is how your text sounds in speech:”} for TTS). For more complex vision-language tasks—such as image captioning and VQA—we adopt an instruction-based approach: the model first generates a textual response, which is then converted into speech using structured prompts like \textit{“The textual caption is `{caption}'. Therefore, the audio caption is:”}.

\section{Experiments}
This section details our experimental setup, the metrics used to assess performance, and the results achieved.

\subsection{Implementation Details}

We use Qwen2.5-1.5B as our backbone LLM and train it with PyTorch. The vision encoder is CLIP-Large-Patch14-336~\cite{radford2021learning}, which produces 256 tokens per image. Speech data is handled by the SpeechTokenizer~\cite{zhang2023speechtokenizer}, which encodes each 1-second segment of audio into 50 discrete tokens. When higher speech fidelity is required, SoundStorm-SpeechTokenizer\footnotemark{} extends this quantization approach to more nuanced tasks. For speech-input settings in image captioning and VQA, we use Melon-TTS to generate spoken questions or prompts, using the same configuration as in the training set, including a mix of speaking speeds and accents. More complementary details are provided in Appendix~\ref{sec:appendix_experiments}.

\footnotetext{https://github.com/ZhangXInFD/soundstorm-speechtokenizer}

\subsection{Metrics}

\paragraph{Text-Output tasks:} For tasks where the model generates text outputs, we use Word Error Rate (WER) for ASR, CIDEr  \cite{vedantam2015cider} for image captioning, and accuracy for VQA. These metrics provide a standardized evaluation framework for assessing performance across text-output tasks.

\paragraph{Speech-output tasks:}
We evaluate speech generation using two methods: We transcribe model-generated speech with Whisper Medium  \cite{radford2023robust} and compare the resulting text to human references (as in text-output tasks).
However, ASR models can exhibit biases that introduce transcription errors, thereby distorting the perceived quality of the generated speech. Therefore, we use WavLM-TDNN\footnotemark to extract speech embeddings from both generated and reference speech and measure their similarity (via cosine similarity). This directly compares acoustic properties without relying on ASR.

\footnotetext{\url{https://github.com/yangdongchao/UniAudio/blob/main/UniAudio/tools/evaluation/compute_similarity_vc.py}}

\subsection{Results}
\begin{table*}[t]
\caption{Performance of Text-Speech Tasks. We evaluate ASR on LibriSpeech  \cite{panayotov2015librispeech} \textit{test-clean} and TTS on VCTK  \cite{Veaux2017CSTRVC}}
\centering
\resizebox{0.7\textwidth}{!}{
\begin{tabular}{l|c|c|cc}
\toprule
\multirow{2}{*}{Model} & \multirow{2}{*}{Backbone} & ASR & \multicolumn{2}{c}{TTS} \\
 &  & LibriSpeech & \multicolumn{2}{c}{VCTK} \\
 &  & WER & WER & Similarity \\
\midrule
Human-level & - & 5.8 & 1.9 & 0.93    \\
\midrule
Wav2vec 2.0  \cite{papineni2002bleu} & - & 2.7 & - &-  \\
Whisper Large V3  \cite{radford2023robust}& - & \textbf{1.8} & - &  -\\
\midrule
VALL-E  \cite{wang2023neural} & - & - & 7.9 & 0.75 \\
VILA  \cite{fu2024vita}& Mixtral 8x7B  \cite{jiang2024mixtral}& 8.1 & - & - \\
AnyGPT-7B & LLaMA-2-7B & 8.5 & 8.5 & \textbf{0.77} \\
MIO-Ins  \cite{wang2024mio}& Yi-6B-Base  \cite{ai2024yi}& 10.3 & \textbf{4.2} & - \\
Qwen2.5-Omni-7B \cite{xu2025qwen2}& Qwen2.5-7B  \cite{ai2024yi}& \textbf{1.8} & - & - \\
\midrule
\textbf{SVLA-2B} & Qwen-1.5B & 10.2 & 21.7 & 0.65 \\
\textbf{SVLA-2B-Text-Ins} & Qwen-1.5B & 8.9 & 11.2 & \textbf{0.72}\\
\bottomrule
\end{tabular}
}
\label{tab:text-speech-performance}
\end{table*}

\subfour{Text-Speech Performance:} 
As shown in Table~\ref{tab:text-speech-performance}, SVLA-2B-Text-Ins outperforms SVLA-2B on both ASR and TTS, reducing ASR WER from 10.2 to 8.9 and TTS WER from 21.7 to 12  .2, while improving similarity from 0.65 to 0.72. Prompting with structured text (e.g., “This is the transcript:”) boosts performance by providing clearer context. While both models lag behind specialized systems like Whisper and Wav2vec 2.0 (WER 2.7), SVLA-2B-Text-Ins is competitive with AnyGPT-7B. The results highlight a trade-off: direct speech generation is more natural, but instruction-based prompts yield greater clarity and accuracy.

\begin{table*}[!t]
\centering
\caption{Comparison of models on Image Captioning and VQA tasks. We evaluate Image Captioning on \textit{COCO-Caption-2014}, \textit{COCO-Caption-2017}, and \textit{Flickr8k} datasets, while VQA performance is assessed on \textit{VQAv2-val}, \textit{OKVQA-test}, \textit{GQA-test}, and \textit{VizWiz} datasets. Modalities are denoted as I (Image), T (Text), and S (Speech). * indicates results on the test-dev set.}
\resizebox{0.98\textwidth}{!}{
\begin{tabular}{l|c|c|ccc|cccc}
\toprule
\multirow{2}{*}{\textbf{Model}} & \multirow{2}{*}{\textbf{Backbone}} & \multirow{2}{*}{\textbf{Input $\rightarrow$ Ouput}} 
& \multicolumn{3}{c|}{\textbf{Image Captioning}} & \multicolumn{4}{c}{\textbf{VQA}}  \\
\cmidrule{4-10}
& & & \textbf{COCO-2014-test} & \textbf{COCO-2017-test} & \textbf{Flickr8k} 
& \textbf{VQAv2-val} & \textbf{OKVQA-test} & \textbf{GQA-test} & \textbf{VizWiz}\\
\midrule
TMT  \cite{kim2024tmt} & - & I$\rightarrow$T & 108.7 & -- & 79.7 &  & -- & -- & -- \\
TMT  \cite{kim2024tmt} & - & I$\rightarrow$S & 78.7 & -- & 55.2 &  & -- & -- & -- \\
\midrule

InstructBLIP  \cite{liu2024visual}& Vicuna-7B  & I+T$\rightarrow$T & -- & 102.2 & 82.2 & -- & 33.9 & -- & 33.4  \\
LLaVA  \cite{liu2024visual}& LLaMA-2-7B& I+T$\rightarrow$T & -- & -- & 82.7 & -- & -- & -- & --  \\
LLaVA-1.5  \cite{liu2024visual}& Vicuna-7B  & I+T$\rightarrow$T & -- & -- & -- & 78.5* & -- & 62.0 & 50.0  \\
AnyGPT-7B  \cite{tang2023any}& LLaMA-2-7B & I+T$\rightarrow$T & 107.5 & -- & -- & -- & -- & -- & -- \\
CoDi  \cite{tang2023anytoany}& - & I+T$\rightarrow$T & 149.9 & -- & -- & -- & -- & -- & -- \\
MIO-Ins  \cite{wang2024mio}& Yi-6B-Base & I+T$\rightarrow$T & 120.4 & -- & -- & 65.5 & 39.9 &  & 53.5  \\
Next-GPT-7B  \cite{wu2024next}& LLama-7B & I+T$\rightarrow$T & 158.3 & 124.9 & 84.5 & 66.7 & 52.1 & -- & 48.4  \\
\midrule

SVLA-2B & Qwen-1.5B &  I+T$\rightarrow$T & 120.0 & 117.8 & 61.4 & 68.7 & 45.4 & 53.3 & 57.7 \\
& Qwen-1.5B &  I+S$\rightarrow$T & 114.5 & 107.0 & 57.7 & 52.9 & 25.1 & 37.7 & 52.3  \\
& Qwen-1.5B & I+T$\rightarrow$S & 2.0 & 2.2 & 1.7 & 4.0 & 4.0 & 3.7 & 0.0 \\
& Qwen-1.5B & I+S$\rightarrow$S & 2.0 & 2.1 & 1.1 & 3.1 & 0.1 & 0.0 & 0.0  \\
\midrule
SVLA-2B-Text-Ins & Qwen-1.5B & I+T$\rightarrow$T & 120.2 & 117.0 & 67.7 & 69.7 & 47.4 & 52.7 & 58.0 \\
& Qwen-1.5B &  I+S$\rightarrow$T & 119.4 & 112.6 & 59.2 & 52.7 & 28.7 & 38.1 & 51.7  \\
& Qwen-1.5B &  I+T$\rightarrow$S & 64.7 & 53.36 & 49.4 & 37.5 & 11.6 & 29.6 & 29.8  \\
& Qwen-1.5B &  I+S$\rightarrow$S & 62.2 & 52.18 & 46.4 & 29.4 & 6.08 & 23.7 & 26.1  \\
\bottomrule
\end{tabular}}

\label{tab:model_performance}
\end{table*}

\subfour{Image Captioning Performance:}
From Table~\ref{tab:model_performance}, instruction tuning in SVLA-2B-Text-Ins leads to minimal change in text-only captioning performance. For instance, on COCO-2014, the CIDEr score improves only slightly from 120.0 (SVLA-2B) to 120.2. However, in the speech captioning setting (I+S→S), instruction tuning results in a substantial gain: SVLA-2B scores only 2.0, while SVLA-2B-Text-Ins reaches 62.2. This highlights the effectiveness of structured prompts (e.g., \textit{``The textual caption is ... Therefore, the audio caption is:''}) in guiding coherent speech generation.
TMT~\cite{kim2024tmt}, which performs I→S generation directly, achieves a CIDEr score of 78.7 on COCO-2014. While upper SVLA-2B-Text-Ins, it operates under a different paradigm—treating each modality independently—whereas SVLA supports unified multimodal reasoning across both text and speech outputs.

\subfour{VQA Performance:}
Table~\ref{tab:model_performance} shows that SVLA-2B-Text-Ins performs competitively in the I+T→T VQA setting, despite using the smaller Qwen-1.5B backbone compared to 7B-scale models. While models like LLaVA-1.5 and Next-GPT-7B achieve strong performance on benchmarks such as VQAv2, OKVQA, and GQA, SVLA-2B-Text-Ins achieves comparable results, with a VQAv2 score of 69.7 and a VizWiz score of 58.0. These results demonstrate that a smaller, instruction-tuned model can rival or even surpass larger alternatives.

However, performance declines in the I+S→T setting, where the question is spoken. For example, SVLA-2B-Text-Ins drops to 52.7 on VQAv2 and 28.7 on OKVQA—approximately 15–20 points lower than in the I+T→T setting. This suggests that the model performs more effectively when the input question is provided in text rather than speech.
Speech output accuracy is generally lower than text output across VQA tasks. In the I+T→S setting, SVLA-2B-Text-Ins achieves 37.5 on VQAv2 and 11.6 on OKVQA, whereas in I+S→S, the scores fall to 29.4 and 6.08, respectively. This drop highlights the difficulty of reasoning directly from speech inputs and generating accurate spoken responses. These results suggest that using text as an intermediate representation enhances semantic alignment, particularly for complex reasoning tasks. By contrast, SVLA-2B completely fails to handle VQA tasks.

\subsection{Ablation Studies}
\subfour{Effect of Accent and Speaking Speed:}

\begin{figure}[htbp]
  \centering
  \begin{subfigure}[b]{0.57\linewidth}
    \includegraphics[width=\linewidth]{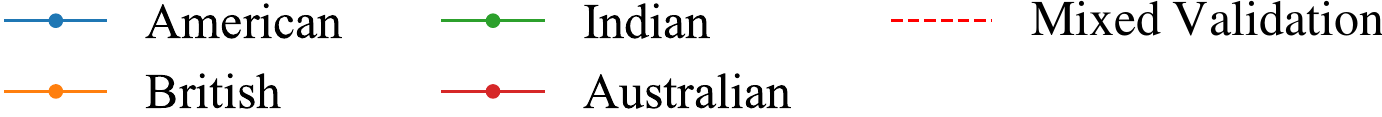}
  \end{subfigure}

  \begin{subfigure}[b]{0.49\linewidth}
    \includegraphics[width=\linewidth]{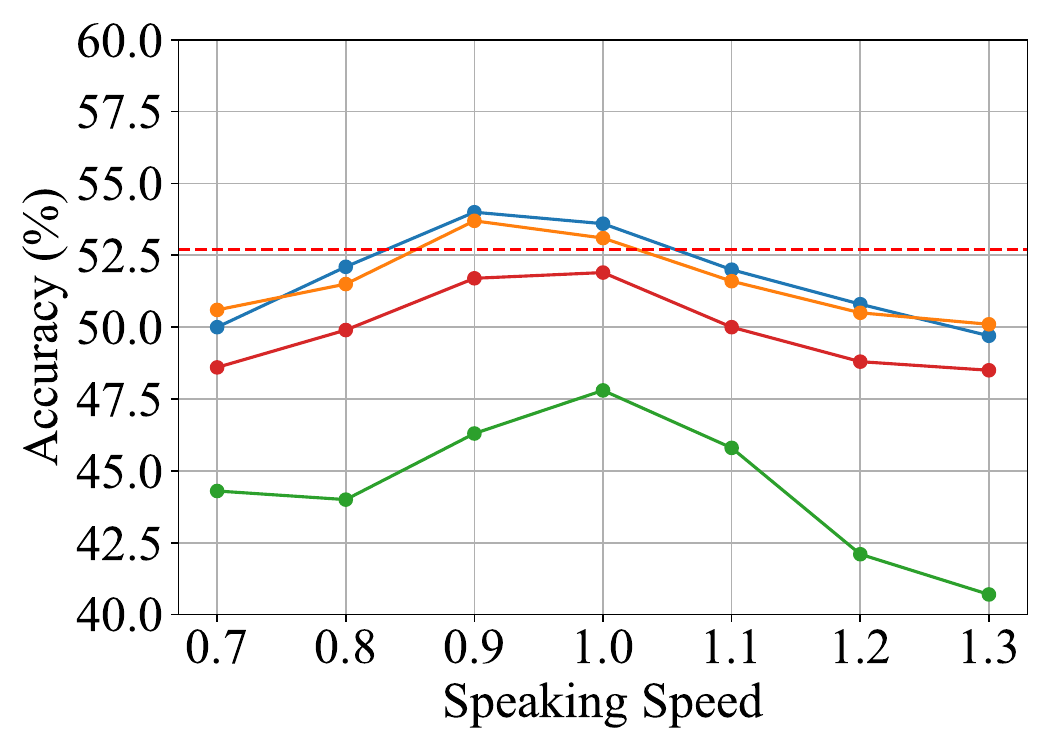}
    \caption{I+T→S}
    \label{fig:vqav2-iss}
  \end{subfigure}
  \hfill
  \begin{subfigure}[b]{0.49\linewidth}
    \includegraphics[width=\linewidth]{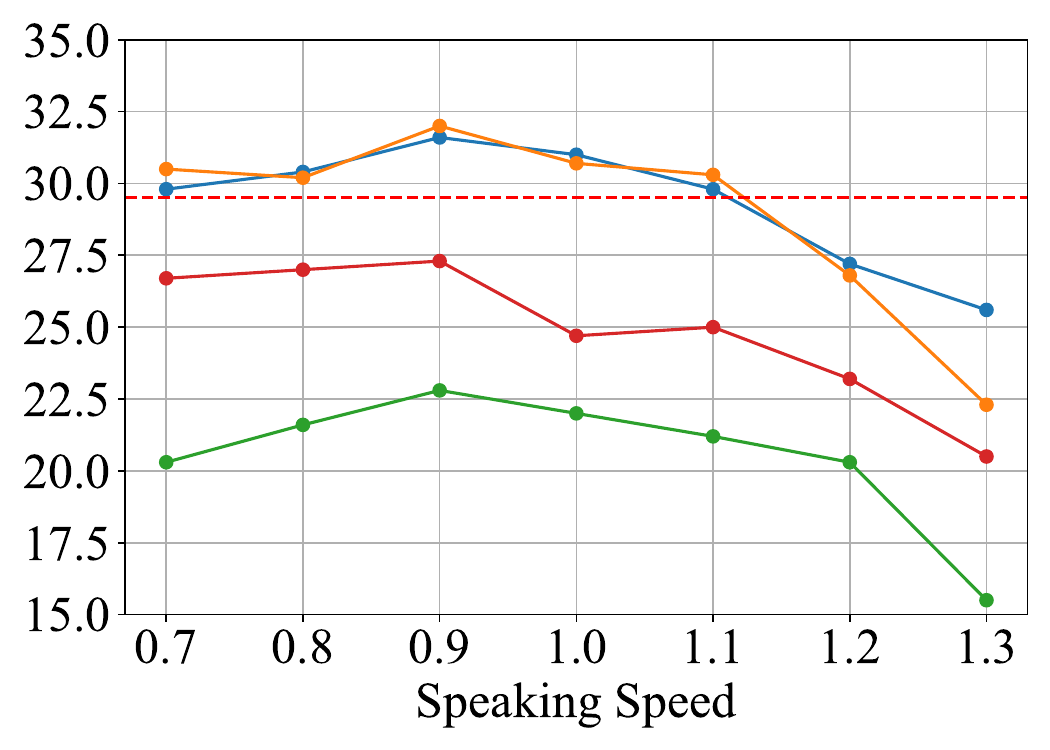}
    \caption{I+S→S}
    \label{fig:vqav2-its}
  \end{subfigure}

  \caption{Impact of accent and speaking speed on VQA performance. Both plots show accuracy on VQAv2-\textit{val}.}
  \label{fig:ablation_speech_variants}
\end{figure}

We evaluate the impact of different accents and speaking speeds on VQA accuracy using the SVLA-2B-Text-Ins model on the VQAv2-\textit{val} set. As show in table \ref{fig:ablation_speech_variants}, in both the I+S→T and I+S→S settings, American and British accents yield higher performance, while Indian and Australian accents result in noticeably lower accuracy. Notably, despite being included in the training data, the Indian accent still underperforms, indicating potential challenges in generalization or speech variability. In terms of speaking speed, the model performs best around the default rate (1.0×), with accuracy dropping at both extremes. The decline is most pronounced at 1.3×, suggesting that faster speech reduces recognition and reasoning quality.

\begin{figure*}[t]
    \centering
    \begin{subfigure}{0.9\textwidth}
        \centering
        \includegraphics[width=\linewidth]{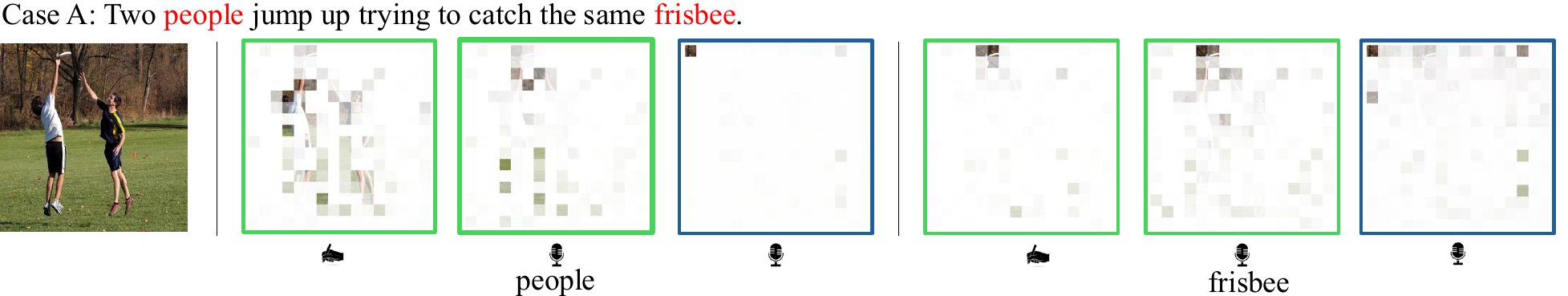}
    \end{subfigure}
    \hfill
    \begin{subfigure}{0.9\textwidth}
        \centering
        \includegraphics[width=\linewidth]{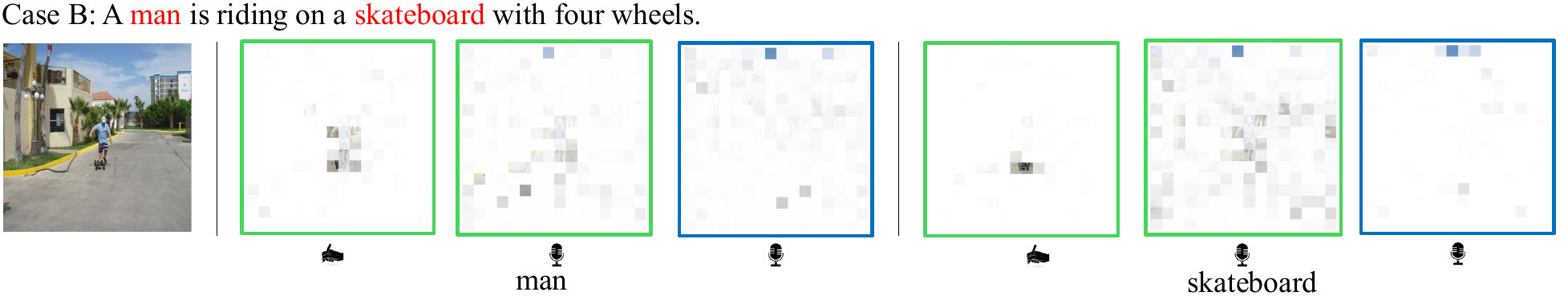}
    \end{subfigure}
    \hfill
    \begin{subfigure}{0.9\textwidth}
        \centering
        \includegraphics[width=\linewidth]{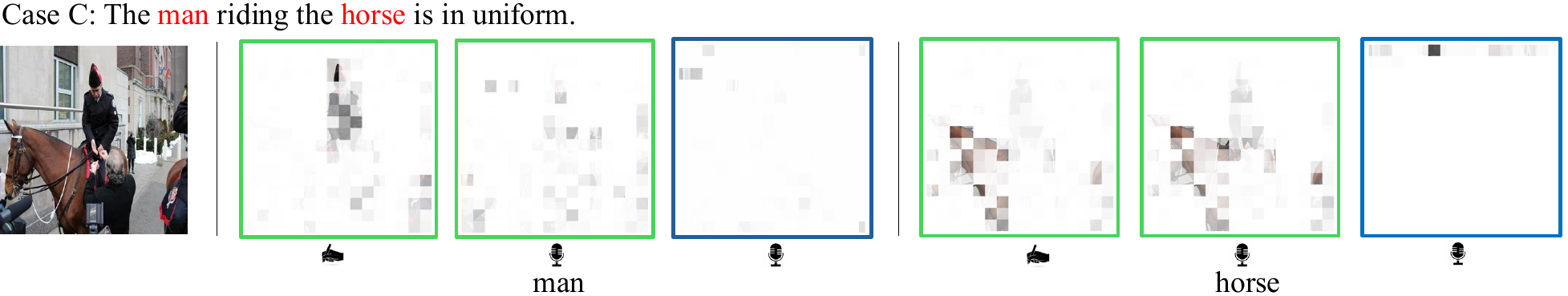}
    \end{subfigure}
    \caption{Visualization of attention maps comparing SVLA-2B's visual grounding accuracy with and without intermediate textual instructions during speech generation.}
    \label{fig:attention}
\end{figure*}

\paragraph{Where Do the Models Look in Images?:}
To examine how different speech generation strategies affect visual grounding, we visualize the model’s attention maps in Figure~\ref{fig:attention}. For each key word, we show three attention maps: the first green map is from the text output of SVLA-2B-Text-Ins, the second green map from its instruction-based speech output; and the blue map from SVLA-2B’s direct speech output.The text output from SVLA-2B-Text-Ins exhibits the most focused and accurate attention, precisely grounding visual entities. Its speech output (green) generally retains meaningful grounding, showing acceptable attention consistency. In contrast, the speech output from SVLA-2B (blue) is often unfocused or misaligned, failing to attend to the relevant image regions.
These results demonstrate that textual instructions play a critical role in guiding the model’s visual attention. Without the intermediate text step, the model lacks semantic anchoring and often fails to locate the correct objects in the image, leading to degraded visual grounding during speech generation.

\begin{figure*}[t]
    \centering
    
    \begin{subfigure}{0.45\textwidth}  
        \centering
        \includegraphics[width=\linewidth]{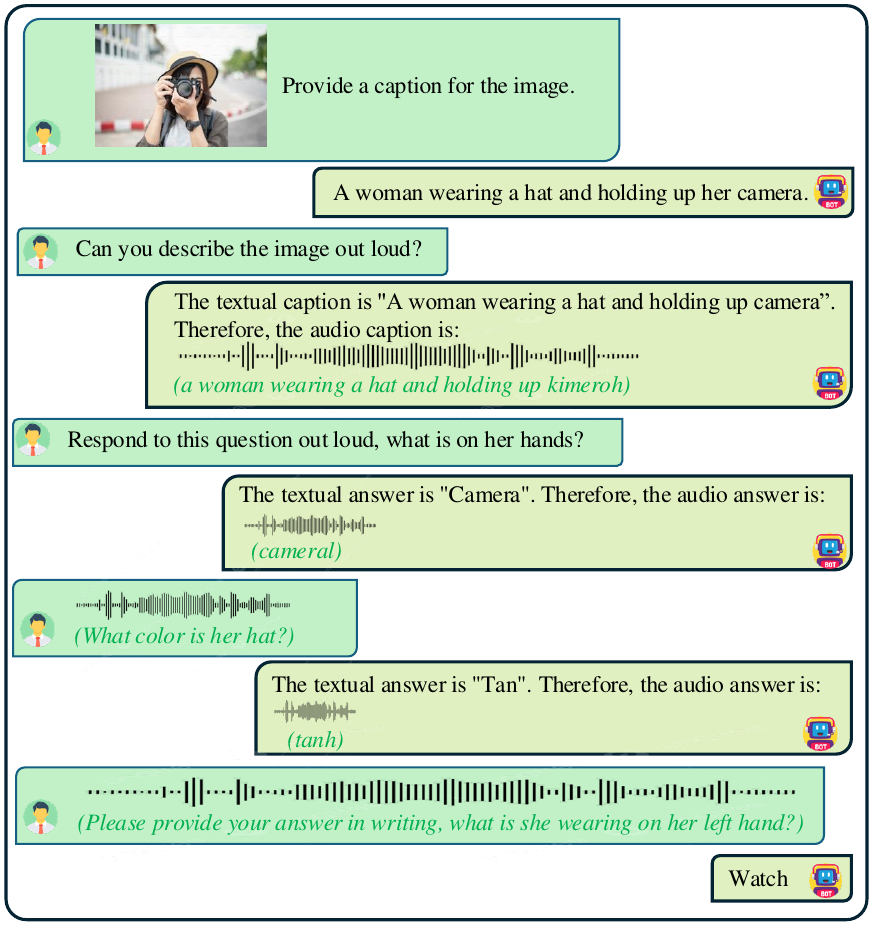}
        \caption{SVLA-2B-Text-Ins}
        \label{fig:text_ins_demo}
    \end{subfigure}
    \hspace{2mm}  
    \begin{subfigure}{0.45\textwidth}  
        \centering
        \includegraphics[width=\linewidth]{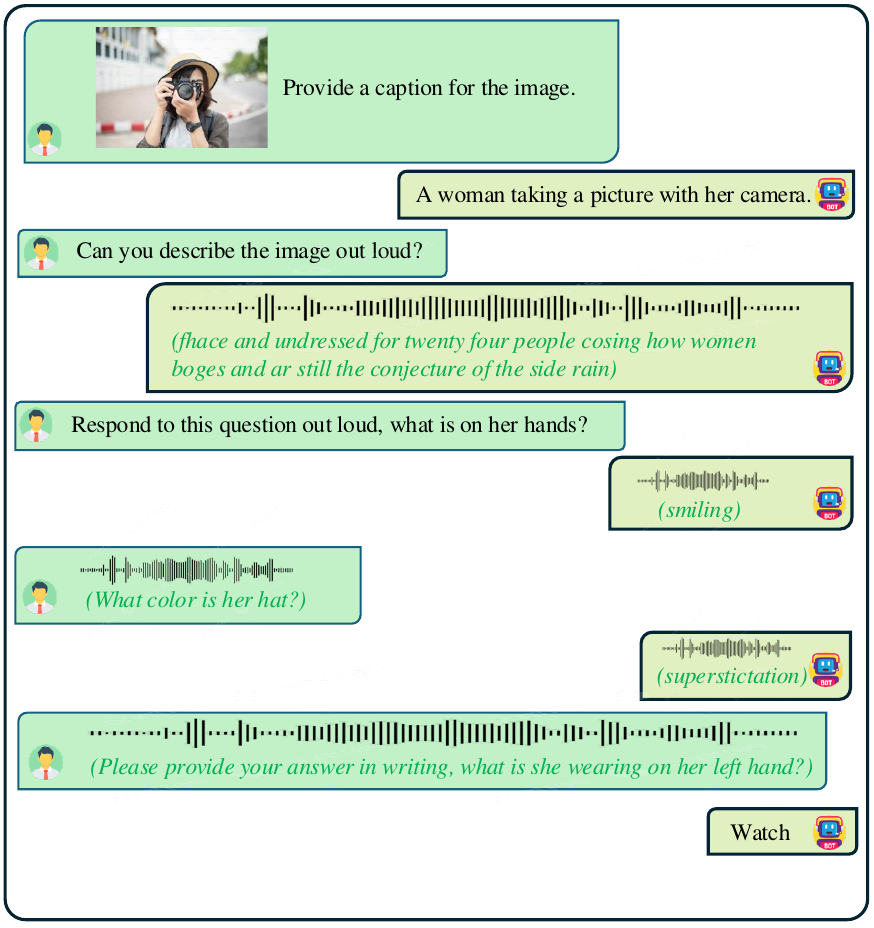}
        \caption{SVLA-2B}
        \label{SVLA-2B}
    \end{subfigure}
    \caption{Comparison of SVLA-2B-Text-Ins and SVLA-2B in Multimodal Image Captioning and VQA Responses.}
    \label{fig:ablation_example}
\end{figure*}


\paragraph{The Limits of ASR-Based Evaluation:} 
ASR-based evaluation falls short in assessing speech output quality, as it often misinterprets minor phonetic variations as errors—even when the spoken response is semantically accurate. As shown in Figure~\ref{fig:text_ins_demo}, the model’s speech output closely matches the intended text, yet ASR transcribes words like \textit{“camera”} as \textit{“camaral”}, or \textit{“tan”} as \textit{“tanh”}. These subtle differences, while perceptually acceptable, are unfairly penalized because ASR prioritizes exact word-level matching over acoustic or semantic similarity.
This limitation underscores the need for more robust evaluation metrics that go beyond transcription accuracy. In particular, speech-based VQA and captioning tasks would benefit from metrics that directly assess the fidelity of the generated speech waveform—capturing both semantic correctness and acoustic naturalness—without relying solely on error-prone intermediate transcriptions.

\section{Ethical Considerations}
We utilize publicly available datasets containing licensed image-text and speech-text pairs. All speech samples are either synthetic or derived from open corpora, explicitly excluding personal or sensitive data. While our datasets incorporate diverse accents and varying speaking rates to enhance representativeness, synthetic speech may still not capture the complete variability inherent in natural human speech. Observed performance disparities across different conditions highlight the necessity of ongoing research in fairness and robustness.

\section{Limitations:}
Our study has several limitations. First, the speech data is generated using a TTS model, which may lack the natural prosody and emotional variation of real human speech. Despite augmentations in accent, speed, and noise, the resulting speech may still be less diverse than natural input. Second, we use the Qwen2.5-1.5B backbone due to resource constraints, which limits model capacity. Third, the speech tokenizer introduces decoding errors, reducing intelligibility even when the underlying text is accurate.

\section{Conclusion}
In this work, we introduce SVLA, a unified Speech-Text-Vision Assistant capable of handling both language tasks (ASR, TTS) and vision-language reasoning tasks (image captioning, VQA). To support the community in building similar models, we also release a large-scale tri-modal dataset encompassing speech, text, and vision. Additionally, we analyze two settings for speech generation: directly producing spoken output and using a text prompt to guide speech synthesis. Our experiments show that while the model performs better with text outputs, speech outputs benefit from an instructive text prompt, yielding more coherent. In future work, we plan to incorporate real human speech, improve speech tokenization quality, and explore larger model backbones to better support nuanced prosody, robustness to speech variability, and high-fidelity speech generation.

\bibliographystyle{unsrt}  
\bibliography{paper_1}

\onecolumn
\appendix
\renewcommand{\thefigure}{\arabic{figure}}
\renewcommand{\thetable}{\arabic{table}}
\setcounter{figure}{0}
\setcounter{table}{0}

\clearpage
\section{Training}
\label{sec:appendix_data_generation}

\subsection{Data Generation}
\begin{table*}[htbp]
\centering
\resizebox{0.8\textwidth}{!}{
\begin{tabular}{llrrrrrr}
\toprule
\rowcolor{blue!30}    Dataset & Task & No. Sample &  Image & Text Tokens & Speech Tokens & Speech Duration (s) & Speech Duration (h) \\
\midrule
\rowcolor{blue!10} \multicolumn{8}{c}{\textbf{Pretrain Dataset - Stage 1}} \\
\midrule
\multicolumn{8}{l}{\footnotesize \textit{Text-Speech tasks}} \\
\midrule
\multirow{2}{*}{Libriheavy} & TTS & 1.0M & 0 & 56.7M & 730M & 14.6M & 4.1K \\
                            & ASR & 1.0M & 0 & 60.0M & 736M & 14.7M & 4.1K \\
\midrule
\rowcolor{red!10} Total& -    &      2.0M &      0 &      1.5M &          7.4B &               29.3M &                8.2K \\
\bottomrule
\rowcolor{blue!10} \multicolumn{8}{c}{\textbf{Pretrain Dataset - Stage 2}} \\
\midrule
\multicolumn{8}{l}{\footnotesize \textit{Text-Speech tasks}} \\
\midrule
\multirow{2}{*}{Libriheavy} & TTS & 3.0M & 0 & 151.6M & 2.2B & 44.2M & 12.3K \\
                            & ASR & 3.0M & 0 & 207.6M & 2.2B & 44.4M & 12.3K \\
\midrule
\multicolumn{8}{l}{\footnotesize \textit{Image-Text-Speech tasks}} \\
\midrule
\multirow{4}{*}{Laion} & IC-TTT & 5.8M & \multirow{4}{*}{5.8M} & 111.1M & 0 & 0 & 0 \\
                       & IC-TTS & 5.8M &                       & 48.4M & 1.1B & 22.0M & 6.1K \\
                       & IC-STT & 5.8M &                       & 63.1M & 1.0B & 20.2M & 5.6K \\
                       & IC-STS & 5.8M &                       & 0 & 1.9B & 38.6M & 10.7K \\
\midrule
\multirow{4}{*}{VG} & VQA-TTT & 1.3M & \multirow{4}{*}{108K} & 12.1M & 0 & 0 & 0 \\
                    & VQA-TTS & 1.3M &                         & 19.1M & 97.0M & 1.9M & 538 \\
                    & VQA-STT & 1.3M &                         & 2.9M & 321.0M & 6.4M & 1.8K \\
                    & VQA-STS & 1.3M &                         & 0 & 249.6M & 182.4M & 1.4K \\

\midrule
\rowcolor{red!10}     Total &    - &      34.3M &  23.5M &      615.9M &          9.1B &               182.8M &               50.8K \\
\bottomrule
\end{tabular}
}
\caption{Statistics of the pretraining dataset}
\label{tab:pretrain_dataset}
\end{table*}

\begin{table*}[htbp]
\centering
\resizebox{0.8\textwidth}{!}{%
\begin{tabular}{llrrrrrr}
\toprule
\rowcolor{blue!30}    Dataset & Task & No. Sample &  Image & Text Tokens & Speech Tokens & Speech Duration (s) & Speech Duration (h) \\
\midrule
\rowcolor{red!10} \multicolumn{8}{l}{\footnotesize \textit{Text-Speech tasks}} \\
\midrule
Librispeech & ASR & 150K & 0 & 5.4M & 89.2M & 1.8M & 496.0 \\
CommandVoice & TTS & 388K & 0 & 4.8M & 101.8M & 2.0M & 559.3 \\
\midrule
\rowcolor{red!10} \multicolumn{8}{l}{\footnotesize \textit{Image-Text-Speech tasks}} \\
\midrule
\multirow{4}{*}{VQA} & VQA-TTT & 84K & \multirow{4}{*}{83K} & 4.7M & 0 & 0 & 0 \\
                    & VQA-TTS & 42K &                       & 3.0M & 13M & 260K & 72 \\
                    & VQA-STT & 42K &                       & 282K & 46.1M & 921K & 256 \\
                    & VQA-STS & 84K &                       & 0 & 86.0M & 17.2M & 476 \\
\midrule
\multirow{4}{*}{A-OKVQA} & VQA-TTT & 50K & \multirow{4}{*}{50K} & 520.0K & 0 & 0 & 0 \\
                        & VQA-TTS & 25K &                         & 333K & 1.4M & 28K & 8 \\
                        & VQA-STT & 25K &                         & 31K & 5M & 102K & 28 \\
                        & VQA-STS & 50K &                         & 0 & 9.5M & 190K & 53 \\
\midrule
\multirow{4}{*}{GQA} & VQA-TTT & 72K & \multirow{4}{*}{72K} & 11.5M & 0 & 0 & 0 \\
                    & VQA-TTS & 36K &                         & 6.7M & 26.8M & 536K & 149 \\
                    & VQA-STT & 36K &                         & 528.5K & 100.5M & 2.0M & 558 \\
                    & VQA-STS & 72K &                         & 0 & 168.0M & 3.4M & 933 \\
\midrule
\multirow{4}{*}{VizWiz} & VQA-TTT & 20K & \multirow{4}{*}{20K} & 780.5K & 0 & 0 & 0 \\
                    & VQA-TTS & 10K &                         & 413K & 883K & 17.7K & 5 \\
                    & VQA-STT & 10K &                         & 28.8K & 5.7M & 114.0K & 31.7 \\
                    & VQA-STS & 20K &                         & 0 & 11.5M & 230.7K & 64.1 \\

\midrule
\multirow{4}{*}{COCO-Caption-2014} & IC-TTT & 414K & \multirow{4}{*}{83K} & 107.0M & 0 & 0 & 0 \\
                        & IC-TTS & 212K &                         & 3.3M & 39.4M & 789K & 219 \\
                        & IC-STT & 212K &                         & 2.4M & 51.3M & 1M & 285 \\
                        & IC-STS & 414K &                         & 0 & 163.7M & 3.3M & 909 \\
\midrule

\rowcolor{red!10}     Total &    - &      2.5M &  308K &      152M &          920M &               33.9M &               5102 \\
\bottomrule
\end{tabular}%
}
\caption{Statistics of the SFT dataset}
\label{tab:sft_dataset}
\end{table*}

Table \ref{tab:pretrain_dataset} provides a detailed breakdown of the pretraining dataset, categorized into two stages: Stage 1, which consists solely of text-speech tasks, and Stage 2, which expands to include image-text-speech tasks. In Stage 1, the dataset comprises 2.0M samples from the Libriheavy corpus for TTS and ASR tasks, yielding a total of 1.5M text tokens, 7.4B speech tokens, and approximately 8.2K hours of speech. Stage 2 significantly scales up the dataset, incorporating Libriheavy (6M samples) alongside multimodal datasets such as Laion and VG, covering image-captioning (IC) and VQA tasks in both text and speech modalities. The total dataset spans 34.3M samples, 615.9M text tokens, 9.1B speech tokens, and 50.8K hours of speech, making it one of the most extensive speech-text-vision corpora for multimodal learning. Notably, the dataset supports a diverse set of multimodal tasks, including IC-TTT, IC-TTS, IC-STT, IC-STS, VQA-TTT, VQA-TTS, VQA-STT, and VQA-STS, ensuring broad coverage across different input-output combinations.

Table \ref{tab:sft_dataset} presents the SFT dataset, covering both text-speech and image-text-speech tasks. The dataset includes 2.5M samples, with 308K image samples, 152M text tokens, and 920M speech tokens, totaling 33.9M seconds (5102 hours) of speech data. The text-speech tasks include 150K ASR samples from Librispeech and 388K TTS samples from CommandVoice, contributing 496 and 559 hours of speech, respectively. In the image-text-speech tasks, various VQA datasets (VQA, A-OKVQA, GQA, VizWiz) and COCO-Caption-2014 support text-based (VQA-TTT, IC-TTT), text-to-speech (VQA-TTS, IC-TTS), speech-to-text (VQA-STT, IC-STT), and speech-to-speech (VQA-STS, IC-STS) tasks. The dataset is diverse and well-balanced, ensuring broad multimodal coverage for fine-tuning models on speech, text, and vision-related tasks.
Figure \ref{fig:asr_prompts}, Figure \ref{fig:tts_prompts}, Figure \ref{fig:caption_speech_prompts}, and Figure \ref{fig:vqa_speech_prompts} show the prompts of tasks used to train the models.

\begin{figure*}
\begin{tcolorbox}[colframe=black!50, colback=black!5, sharp corners]

\textbf{ASR (Automatic Speech Recognition) Prompts:}
\begin{itemize}
    \item ``Please convert this audio to text: "
    \item ``Transcribe the following audio file, please: "
    \item ``Can you convert this speech to text? "
    \item ``Generate text from this audio recording: "
    \item ``Please write out what’s being said in this audio: "
    \item ``Turn this voice recording into text, please: "
    \item ``Please create a transcript of this audio: "
    \item ``Can you transcribe this audio? "
    \item ``Convert this spoken content into written text: "
    \item ``Please extract text from this speech: "
    \item ``Transcribe the spoken words in this audio file: "
    \item ``Create a written version of this audio: "
    \item ``Convert the spoken words to text: "
    \item ``Please generate a transcript from this recording: "
    \item ``Transform the audio into a text document: "
\end{itemize}

\textbf{Example Usage:}
\begin{verbatim}
"Please convert this audio to text: <speech_start>{speech_tokens}<speech_end>"
\end{verbatim}

\end{tcolorbox}  
\caption{Prompts for ASR Tasks.}
\label{fig:asr_prompts}
\end{figure*}

\begin{figure*}
\begin{tcolorbox}[colframe=black!50, colback=black!5, sharp corners]

\textbf{TTS (Text-to-Speech) Prompts:}
\begin{itemize}
    \item ``Please convert this text to speech: "
    \item ``Turn this text into audio, please: "
    \item ``Generate speech from this text: "
    \item ``Please speak out this text: "
    \item ``Convert these words to speech, please: "
    \item ``Please make an audio version of this text: "
    \item ``Can you read this text aloud: "
    \item ``Transform this text into speech: "
    \item ``Please give a voice to this text: "
    \item ``Read out this text, please: "
    \item ``Create a spoken version of this text: "
    \item ``Convert the written text to speech: "
    \item ``Turn the following words into sound: "
    \item ``Provide an audio rendition of this text: "
    \item ``Generate an audio file of these words: "
\end{itemize}

\textbf{Example Usage:}
\begin{verbatim}
"Turn this text into audio: {transcript}"
\end{verbatim}

\end{tcolorbox}  
\caption{Prompts for TTS Tasks.}
\label{fig:tts_prompts}
\end{figure*}

\begin{figure*}
\begin{tcolorbox}[colframe=black!50, colback=black!5, sharp corners]

\textbf{Captioning Prompts:}

\textbf{IC\_TTT (Text-to-Text) and Caption\_STS (Speech-to-Speech):}
\begin{itemize}
    \item ``What do you see in the image?"
    \item ``Explain what is shown in the picture."
    \item ``Provide a caption for the image."
    \item ``Describe the objects or people in the image."
    \item ``Tell what is happening in the picture."
    \item ``Explain the scene in the image."
    \item ``Describe the setting of the image."
    \item ``List the elements in the picture."
    \item ``What is the main focus of the image?"

\end{itemize}

\textbf{Example Usage:}
\begin{verbatim}
"<image>\nWhat do you see in the image?"
\end{verbatim}
\noindent\rule{0.98\linewidth}{0.4pt}
\textbf{IC\_TTS (Text-to-Speech):}
\begin{itemize}
    \item ``Can you describe the image out loud?"
    \item ``Read the description of the image aloud."
    \item ``Turn the image caption into spoken words."
    \item ``Provide a spoken description of the picture."
    \item ``Talk about what is shown in the image."
    \item ``Can you vocalize the image details?"
    \item ``Use speech to explain the image."
    \item ``Read the text-based description of the image."
    \item ``Convert the image details into speech."

\end{itemize}

\textbf{Example Usage:}
\begin{verbatim}
"<image>\nCan you describe the image out loud?"
\end{verbatim}
\noindent\rule{0.98\linewidth}{0.4pt}
\textbf{IC\_STT (Speech-to-Text):}
\begin{itemize}
    \item ``Write down what you see in the image."
    \item ``Can you write a detailed description of the picture?"
    \item ``Write a summary of the scene in the image."
    \item ``Write down the elements present in the picture."
    \item ``Write down the key features visible in the image."
    \item ``In text, explain what is happening in the picture."
    \item ``Write a visual interpretation of the image."

\end{itemize}

\textbf{Example Usage:}
\begin{verbatim}
"<image>\n Write down what you see in the image."
\end{verbatim}

\end{tcolorbox}  
\caption{Prompts for Image Captioning Tasks.}
\label{fig:caption_speech_prompts}
\end{figure*}

\begin{figure*}
\begin{tcolorbox}[colframe=black!50, colback=black!5, sharp corners]

\textbf{VQA (Visual Question Answering) Prompts:}

\textbf{VQA\_TTT (Text-to-Text) and VQA\_STS (Speech-to-Speech):}

\textbf{Example Usage:}
\begin{verbatim}
"<image>\n{question}\nAnswer the question using a single word or phrase."
\end{verbatim}
\noindent\rule{0.70\linewidth}{0.4pt}
\textbf{VQA\_STT (Speech-Text-to):}
\begin{itemize}
    \item ``Please provide your answer in writing."
    \item ``Respond to the question with a written explanation."
    \item ``Answer this question using text."
    \item ``Type your response to the question."
    \item ``Write down your answer clearly."
    \item ``Provide a detailed answer in text format."
    \item ``Explain your response in written form."
    \item ``Answer the question by typing a full response."
    \item ``Give your answer in written words."
    \item ``Respond with a written explanation and elaborate where necessary."
    \item ``Use text to explain your answer."
    \item ``Please write your answer instead of speaking it."
    \item ``Type out your answer using complete sentences."
    \item ``Deliver your response in text form."
    \item ``Provide a clear and concise written answer."
\end{itemize}

\textbf{Example Usage:}
\begin{verbatim}
"<image>\nPlease provide your answer in writing.\n{question}\nAnswer the quest
-ion using a single word or phrase."
\end{verbatim}

\noindent\rule{0.98\linewidth}{0.4pt}
\textbf{VQA\_TTS (Text-to-Speech):}
\begin{itemize}
    \item `Respond to this question out loud."
    \item `Please give your answer verbally."
    \item `Provide a spoken response to the question."
    \item `Answer this question using speech."
    \item `Explain your response in spoken form."
    \item `Say your answer instead of writing it down."
    \item `Provide a verbal explanation for the question."
    \item `Speak your answer clearly."
    \item `Give a detailed spoken answer to this question."
    \item `Respond verbally and expand on your answer."
    \item `Use spoken words to explain the answer."
    \item `Speak your response rather than typing it."
    \item `Answer the question aloud with full sentences."
    \item `Deliver your response in speech format."
\end{itemize}

\textbf{Example Usage:}
\begin{verbatim}
"<image>\nRespond to this question out loud.\n{question}\nAnswer the quest
-ion using a single word or phrase."
\end{verbatim}

\end{tcolorbox}  
\caption{Prompts for VQA Tasks.}
\label{fig:vqa_speech_prompts}
\end{figure*}
\subsection{Training Strategy}
\label{subsec:training_stages}
Our model is trained in three sequential phases, progressively increasing multimodal complexity to enhance stability, efficiency, and modality integration.

\paragraph{Stage 1 - Text-Speech Pre-Training} The model learns text-speech alignment through ASR and TTS on large-scale paired datasets, establishing a strong linguistic foundation before incorporating vision.

\paragraph{Stage 2 - Vision-Text-Speech Pre-Training} Training expands to include vision-based tasks like VQA and image captioning alongside ASR and TTS, enabling the model to unify vision, text, and speech representations.

\paragraph{Stage 3 - Supervised Fine-Tuning} The model is refined through supervised fine-tuning on benchmark-aligned datasets, focusing on multi-turn conversations, history retention, and multimodal reasoning.

\clearpage
\section{ Implementation Details}
\label{sec:appendix_experiments}

\begin{table*}[t]
    \centering
    \caption{Implementation Details for Model, Pre-Training, and Supervised Fine-Tuning}
    \label{tab:implementation}
    \resizebox{0.48\textwidth}{!}{%
    \renewcommand{\arraystretch}{1.1}
    \begin{tabular}{l|c|c}
    \hline
    \multicolumn{3}{c}{\textbf{Model and Hardware}} \\
    \hline
    \textbf{LLM} & \multicolumn{2}{c}{Qwen2.5-1.5B} \\
    \textbf{Vision Encoder} & \multicolumn{2}{c}{openai/clip-vit-large-patch14-336} \\
    \textbf{Speech Encoder} & \multicolumn{2}{c}{SpeechTokenizer (1s \(\to\) 50 tokens),} \\
    \textbf{Speech Decoder} & \multicolumn{2}{c}{SoundStorm-SpeechTokenizer} \\
    \textbf{Frameworks} & \multicolumn{2}{c}{PyTorch, DeepSpeed} \\
    \hline
    \multicolumn{3}{c}{\textbf{Training Configurations}} \\
    \hline
    & \textbf{Pre-Training} & \textbf{Supervised Fine-Tuning} \\
    \hline
    \textbf{Batch Size} & 32 & 8 \\
    \textbf{Epochs} & 1 & 2 \\
    \textbf{Optimizer} & Adam & Adam \\
    \textbf{Learning Rate} & \(2 \times 10^{-5}\) & \(1 \times 10^{-5}\) \\
    \textbf{Warmup Ratio} & 0.03 & 0.02 \\
    \textbf{LR Scheduler} & Cosine & Cosine \\
    \textbf{Training Steps} & 1.2M & 600K \\
    \textbf{Training Epochs} & 1 & 2 \\
    \textbf{Maximum Tokens} & 4096 & 4096 \\
    \textbf{GPUs} & 8 x NVIDIA H100 & 4 x NVIDIA H100 \\
    \textbf{DeepSpeed Config} & \multicolumn{2}{c}{\texttt{zero2}} \\
    \hline
    \end{tabular}%
    }
\end{table*}

\begin{figure*}[h]
    \centering
    \begin{subfigure}{0.48\textwidth}
        \centering
        \includegraphics[width=\linewidth]{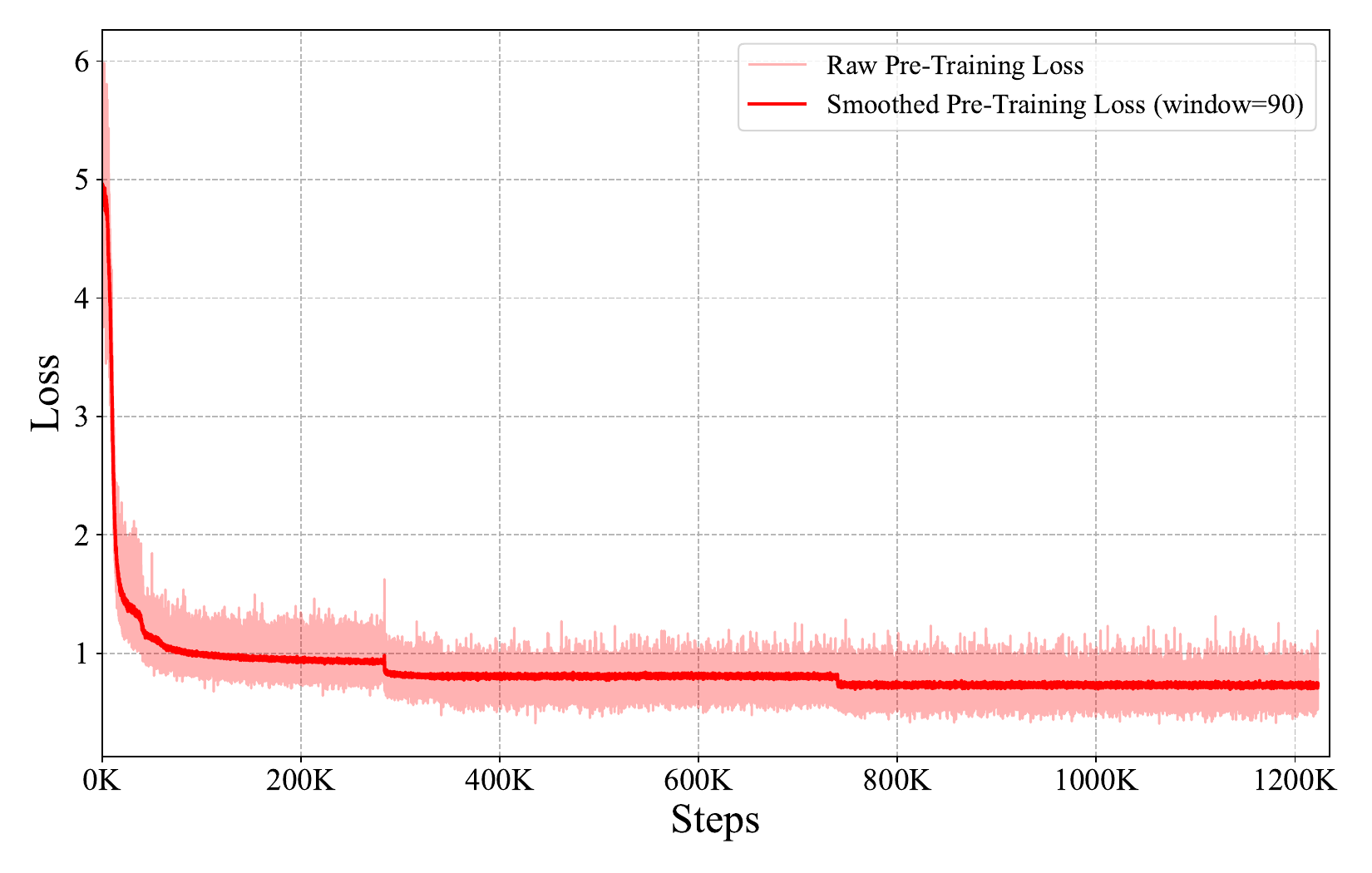}
        \caption{Pre-training Loss}
        \label{fig:pretraining}
    \end{subfigure}
    \hfill
    \begin{subfigure}{0.48\textwidth}
        \centering
        \includegraphics[width=\linewidth]{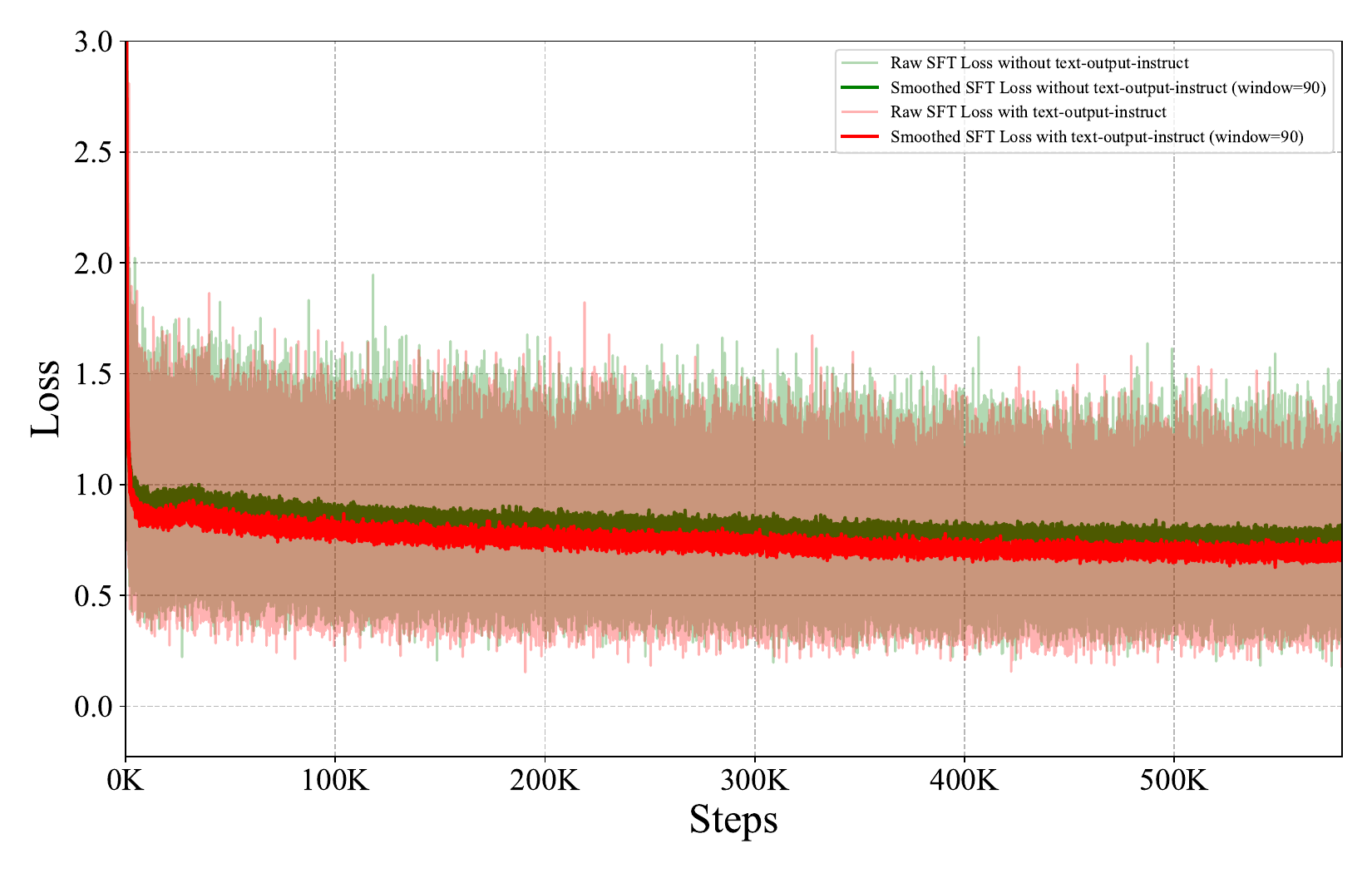}
        \caption{SFT loss}
        \label{fig:sft_loss}
    \end{subfigure}
    \caption{Comparison of Pre-training and SFT Loss Curves}
    \label{fig:loss}
\end{figure*}

Figure~\ref{fig:loss} compares the loss curves for pre-training and supervised fine-tuning (SFT). Figure \ref{fig:pretraining} plot shows the pre-training loss, where the raw loss exhibits high variance but progressively decreases, stabilizing after around 400K steps. The smoothed loss curve (window=90) highlights a consistent downward trend, indicating stable convergence. The \ref{fig:sft_loss} illustrates the SFT loss, comparing models with and without text-output instructions. While both configurations show a decreasing trend, the model trained with text-output instructions (red) achieves consistently lower loss than the version without instructions (green), suggesting that structured textual guidance improves fine-tuning efficiency and convergence. The variance in SFT loss remains higher compared to pre-training, reflecting the increased complexity of supervised instruction tuning.

\begin{figure*}[h]
    \centering
    \begin{subfigure}{0.32\textwidth}
        \centering
    \end{subfigure}
    \begin{subfigure}{0.32\textwidth}
        \centering
        \includegraphics[width=\linewidth]{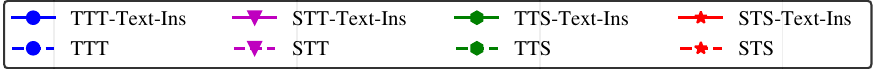}
    \end{subfigure}
    \begin{subfigure}{0.32\textwidth}
        \centering
    \end{subfigure}

    \begin{subfigure}{0.33\textwidth}
        \centering
        \includegraphics[width=\linewidth]{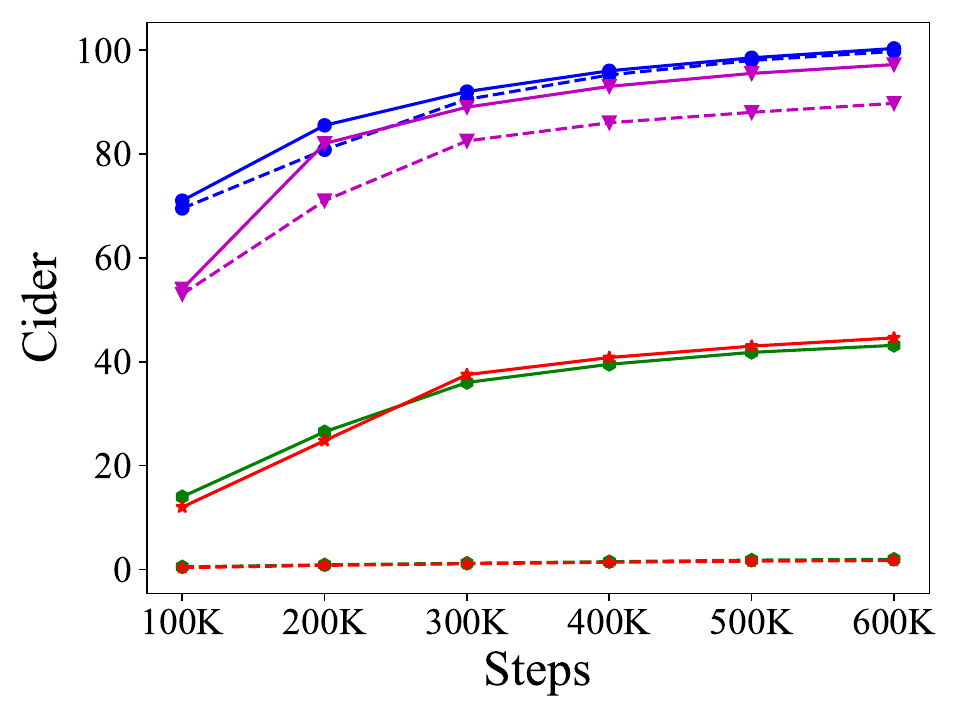}
        \caption{Image Caption Performance}
        \label{fig:caption-performance}
    \end{subfigure}
    \hfill
    \begin{subfigure}{0.33\textwidth}
        \centering
        \includegraphics[width=\linewidth]{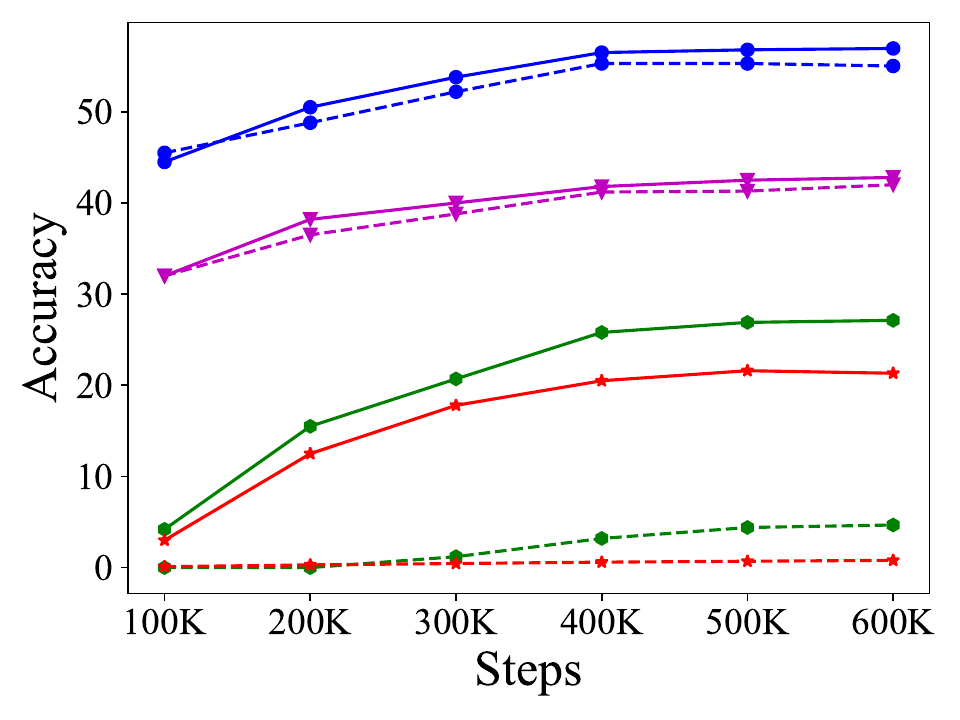}
        \caption{VQA Performance}
        \label{fig:vqa-performance}
    \end{subfigure}
    \hfill
    \begin{subfigure}{0.33\textwidth}
        \centering
        \includegraphics[width=\linewidth]{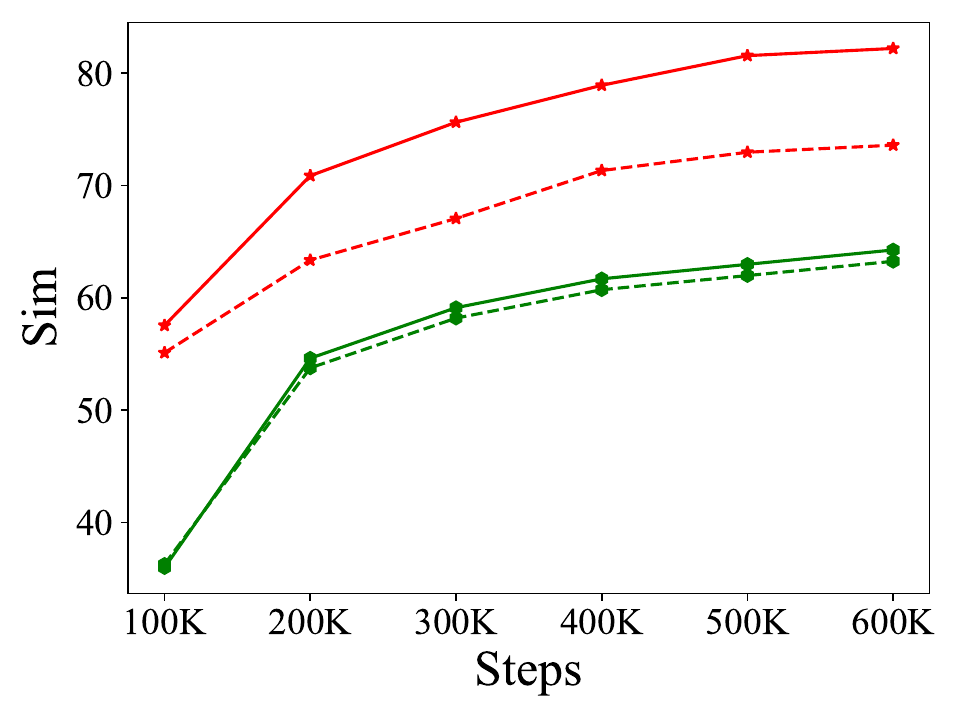}
        \caption{Overall Audio Similarity Performance}
        \label{fig:audio-similarity}
    \end{subfigure}

    \begin{subfigure}{0.33\textwidth}
        \centering
        \includegraphics[width=\linewidth]{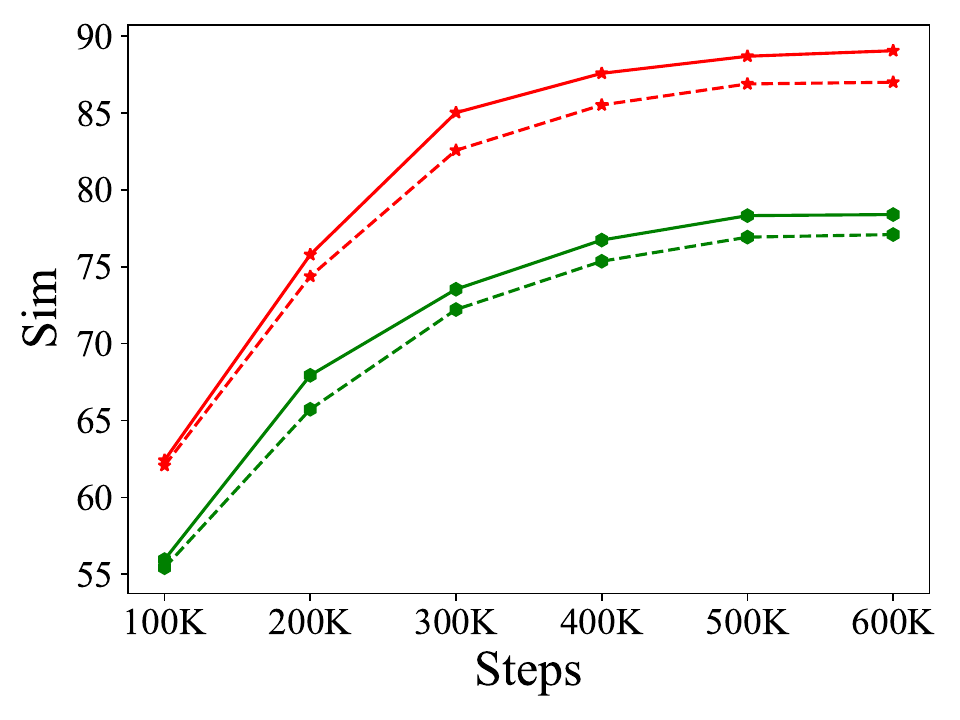}
        \caption{Speech Image Caption Similarity}
        \label{fig:caption-sim}
    \end{subfigure}
    \hfill
    \begin{subfigure}{0.33\textwidth}
        \centering
        \includegraphics[width=\linewidth]{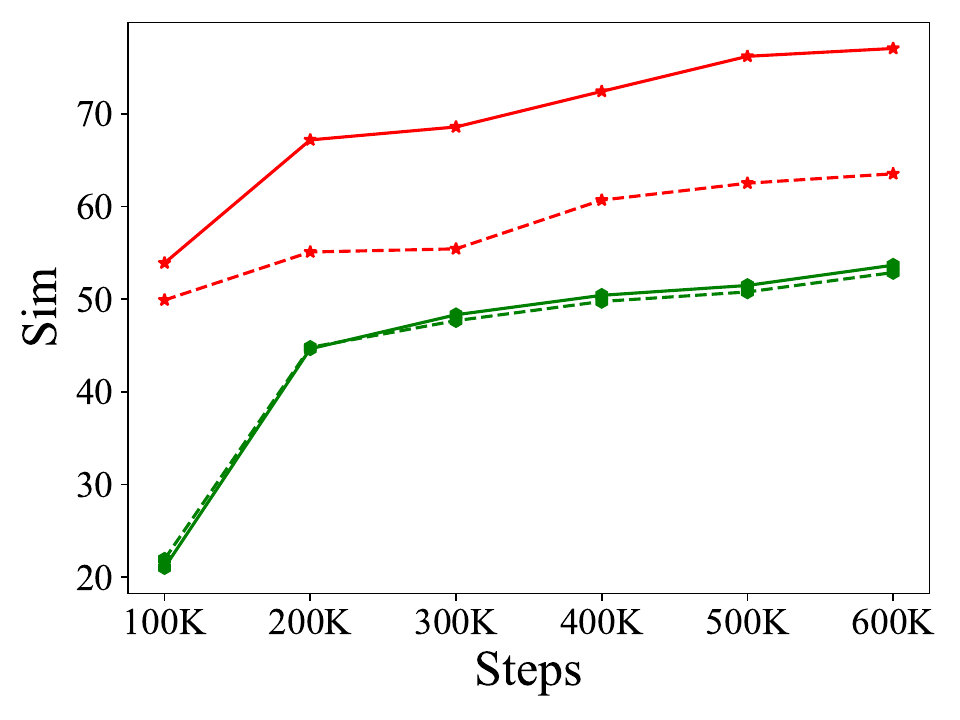}
        \caption{Speech VQA Similarity}
        \label{fig:vqa-sim}
    \end{subfigure}
    \hfill
    \begin{subfigure}{0.33\textwidth}
        \centering
        \includegraphics[width=\linewidth]{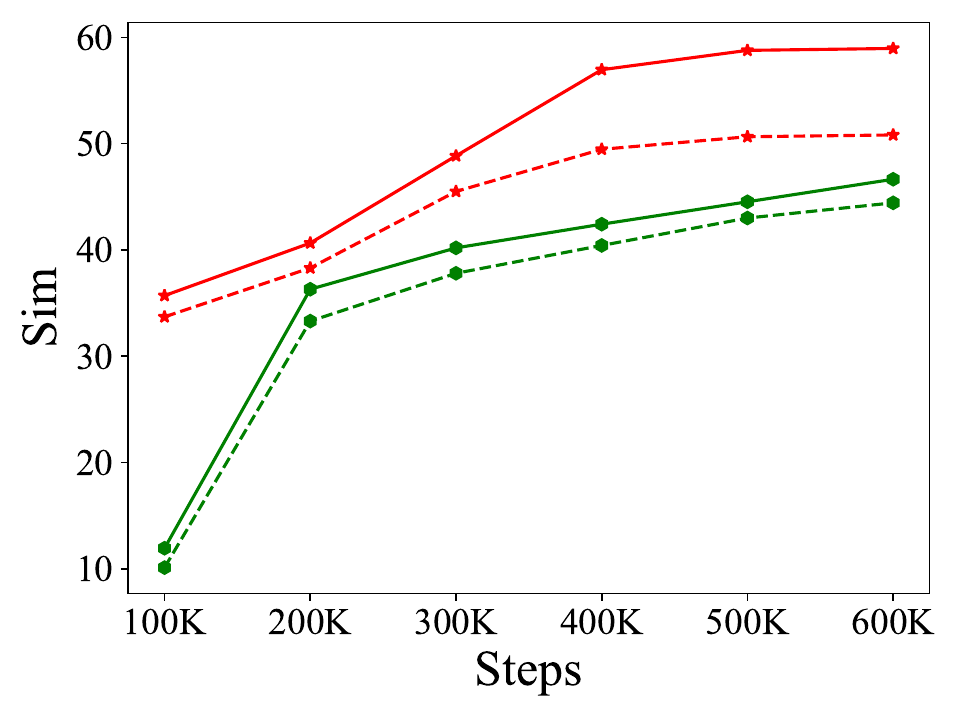}
        \caption{ Weighted Speech VQA Similarity}
        \label{fig:vqa-weighted-sim}
    \end{subfigure}
    \caption{Performance trends of image captioning and  VQA tasks over training steps.}
    \label{figure:step_performance}
\end{figure*}

According to Figure~\ref{figure:step_performance}, the performance for both text and speech outputs consistently improves with additional training steps, but notable differences emerge between modalities. In text-based evaluations (Figures \ref{fig:caption-performance}–\ref{fig:vqa-performance}), accuracy increases quickly and stabilizes after approximately 400K steps, with instruction-tuned models maintaining a moderate advantage over non-instruction models. In contrast, speech-based outputs (Figures \ref{fig:audio-similarity}–\ref{fig:vqa-weighted-sim}), evaluated via acoustic similarity, show a more pronounced improvement from instruction tuning across training steps. The performance gap between instruction-tuned and non-instruction models remains significant throughout, highlighting the critical role of instructions in enhancing speech quality and consistency. Overall, while text outputs rapidly reach a performance plateau, speech outputs benefit more substantially from extended training and explicit instruction tuning.

\clearpage

\section{ Evaluation Details}
\label{sec:appendix_evaluation}

\begin{table}[ht]
    \centering
    \begin{tabular}{l|c c c c}
        \toprule
        \textbf{Generated Modality} & \textbf{Text} & \textbf{Speech} \\
        \midrule
        Beam size & 5 & 1 \\
        Top-P & - & 0.7 \\
        maximum & 256 & 1024 \\
        Repetition Penalty & 1.0 & 1.3\\
        \bottomrule
    \end{tabular}
    \caption{Comparison of Text and Speech Generation Settings in evaluation.}
    \label{tab:decoding_strategies}
\end{table}

\begin{figure*}[ht]
\begin{tcolorbox}[colframe=black!50, colback=black!5, sharp corners]

\textbf{ASR Prompt:}
\begin{verbatim}
"Please convert this audio to text: <speech_start>{speech tokens}<speech_end>."
\end{verbatim}

\textbf{TTS Prompt:}
\begin{verbatim}
"Please convert this text to speech: {transcript}."
\end{verbatim}

\textbf{IC\_TTT and IC\_STS Prompt:}
\begin{verbatim}
"<image>\nrovide a caption for the image."
\end{verbatim}

\textbf{IC\_TTS Prompts:}
\begin{verbatim}
"<image>\nTurn the image caption into spoken words"
\end{verbatim}

\textbf{IC\_STT Prompts:}
\begin{verbatim}
"<image>\nWrite down what you see in the image."
\end{verbatim}

\textbf{VQA\_TTT and VQA\_STS Prompt:}
\begin{verbatim}
"<image>\n{question}\nAnswer the question using a single word or phrase."
\end{verbatim}

\textbf{VQA\_TTS Prompts:}
\begin{verbatim}
"<image>\nAnswer this question using speech.{question}\nAnswer the question 
using a single word or phrase."
\end{verbatim}

\textbf{VQA\_STT Prompts:}
\begin{verbatim}
"<image>\nAnswer this question using text.{question}\nAnswer the question 
using a single word or phrase."
\end{verbatim}

\end{tcolorbox}  
\caption{Evaluation Prompts for different tasks.}
\label{fig:evaluating_prompts}
\end{figure*}

\clearpage

\section{Examples}
\begin{figure*}[!ht]
    \centering
    
    \begin{subfigure}{0.45\textwidth}  
        \centering
        \includegraphics[width=\linewidth]{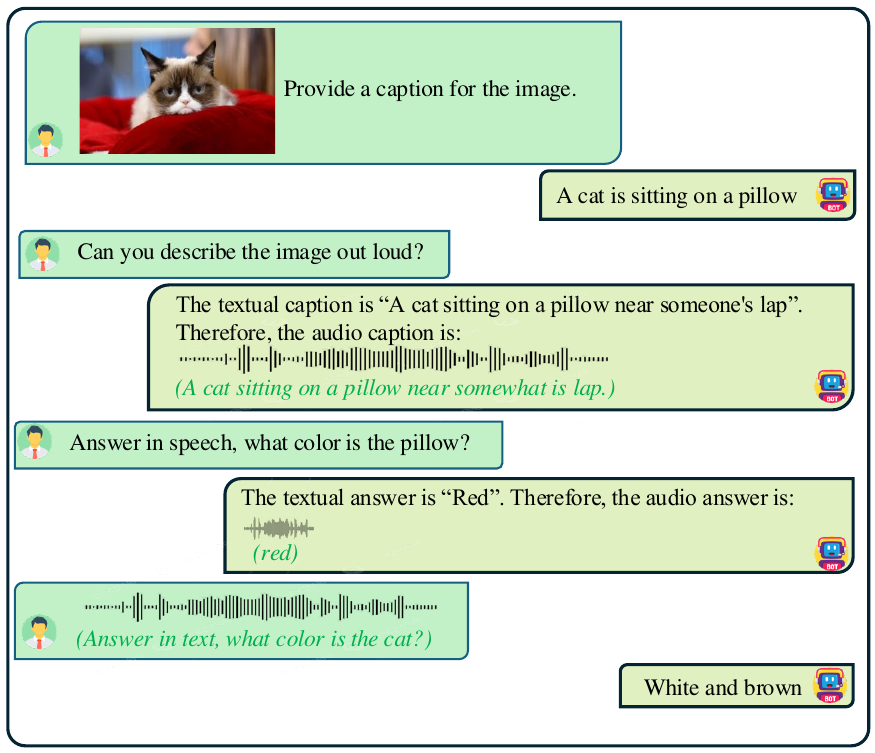}
        \caption{SVLA-2B-Text-Ins}
    \end{subfigure}
    \hspace{2mm}  
    \begin{subfigure}{0.45\textwidth}  
        \centering
        \includegraphics[width=\linewidth]{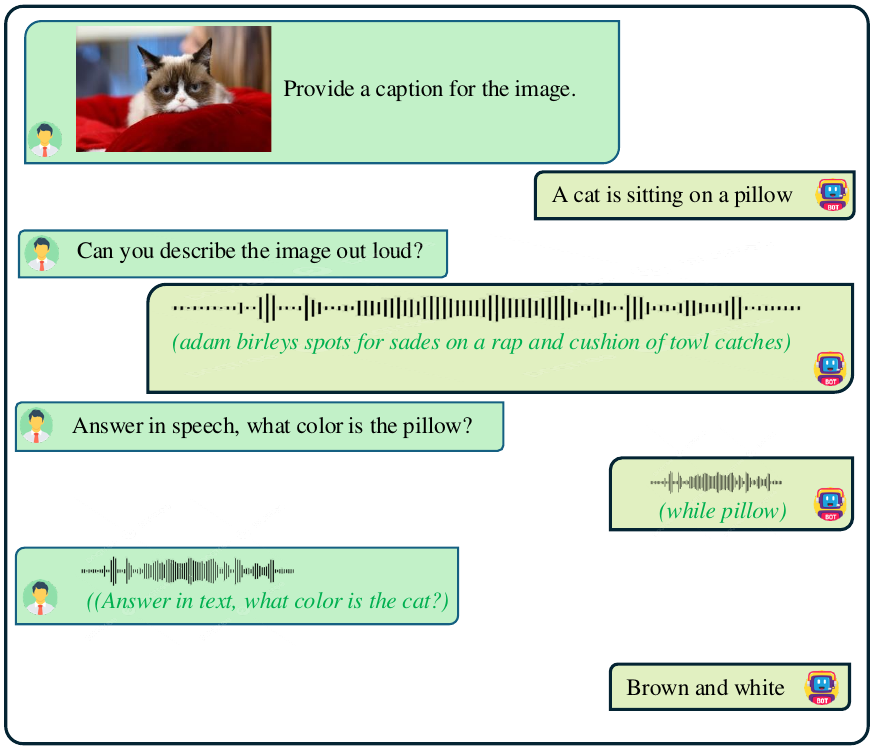}
        \caption{SVLA-2B}
    \end{subfigure}
    
    \caption{Comparison of SVLA-2B-Text-Ins and SVLA-2B in Multimodal Image Captioning and VQA Responses (Example 2).}
\end{figure*}

\begin{figure*}[!ht]
    \centering
    
    \begin{subfigure}{0.45\textwidth}  
        \centering
        \includegraphics[width=\linewidth]{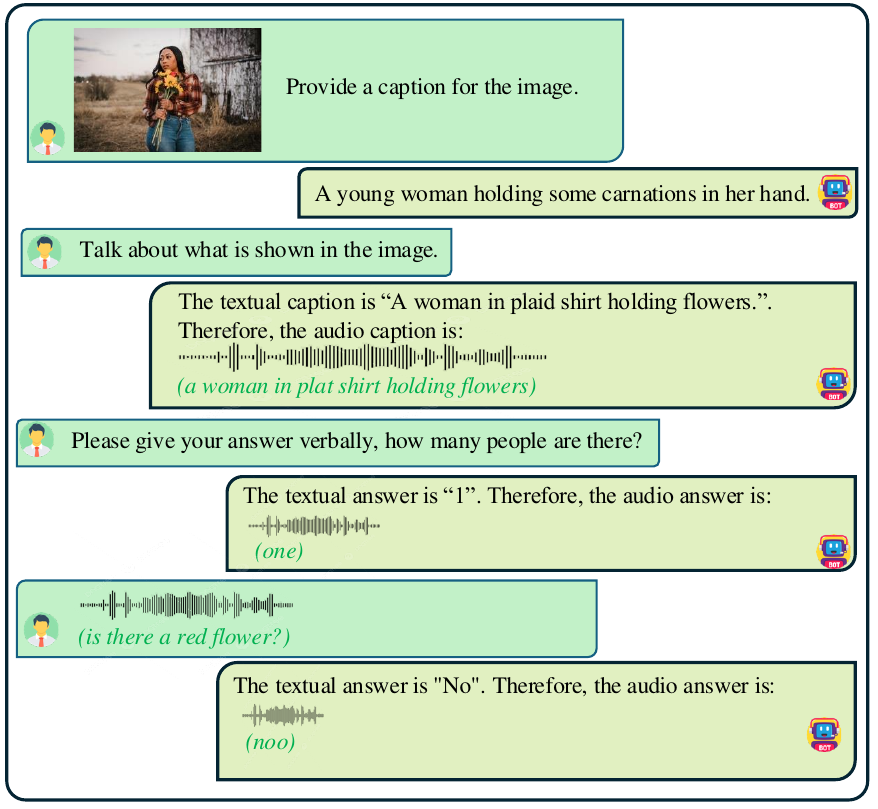}
        \caption{SVLA-2B-Text-Ins}
    \end{subfigure}
    \hspace{2mm}  
    \begin{subfigure}{0.45\textwidth}  
        \centering
        \includegraphics[width=\linewidth]{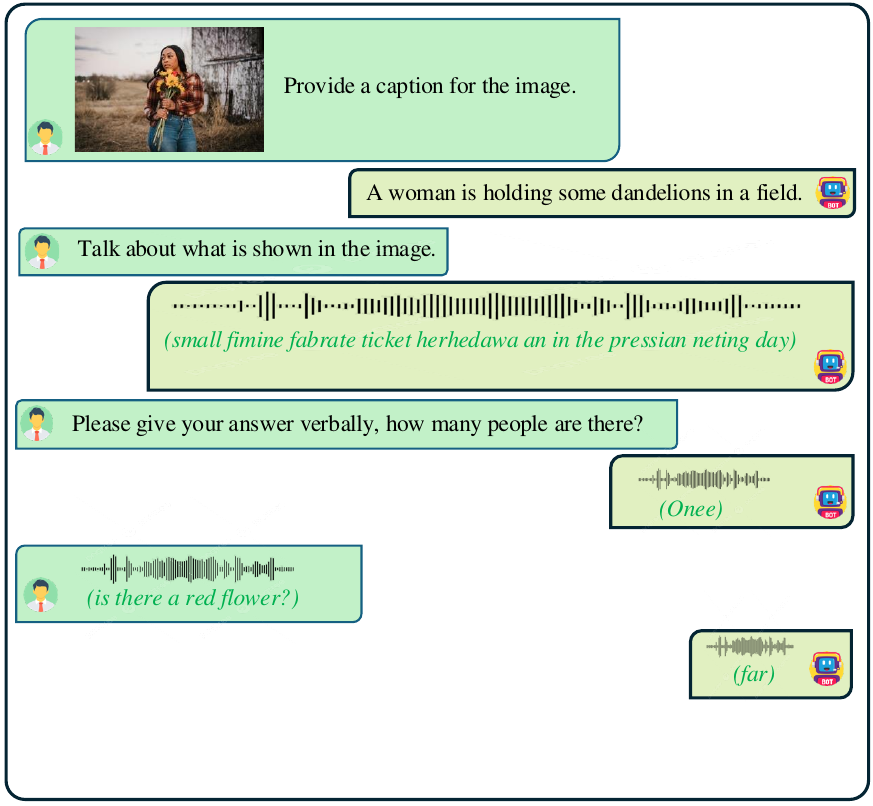}
        \caption{SVLA-2B}
    \end{subfigure}
    
    \caption{Comparison of SVLA-2B-Text-Ins and SVLA-2B in Multimodal Image Captioning and VQA Responses (Example 3).}
\end{figure*}

\end{document}